\documentclass[%
reprint,
amsmath,amssymb,
aps,
floatfix,
]{revtex4-2}
\pdfoutput=1

\usepackage{graphicx}
\usepackage{dcolumn}
\usepackage{bm}
\usepackage{xcolor}
\usepackage{comment}

\usepackage{xr-hyper}
\usepackage{lineno}

\newcommand{\be}{\begin{equation}}
\newcommand{\ee}{\end{equation}}
\usepackage{xspace}
\newcommand{\ket}[1]{\ensuremath{|#1\rangle}\xspace}
\newcommand{\bra}[1]{\ensuremath{\langle #1|}\xspace}
\newcommand{\sprod}[2]{\ensuremath{\langle #1|#2\rangle}\xspace}

\begin{document}
	
	
	\preprint{APS/123-QED}
	
	\title{Generalized Time-Reversal for Pulse Control in Diffusive Media}
	
	\author{Rohin E. McIntosh$^1$}
	\author{Arthur Goetschy$^2$}
	\author{Alexey~Yamilov$^3$}
	\author{Nicholas Bender$^4$}
	\author{Hasan Y\i lmaz$^{5,6}$}
	\author{Hui Cao$^{7,*}$}
	\affiliation{
		$^{1}$Department of Physics, Yale University, New Haven, Connecticut 06520, USA\\
		$^{2}$  Institut  Langevin, ESPCI  Paris, PSL  University,  CNRS, Paris,  France \\
		$^{3}$ Physics Department, Missouri University of Science \& Technology, Rolla, Missouri \\
		$^{4}$  School of Applied and Engineering Physics, Cornell University, Ithaca, New York, USA \\
		$^{5}$ Institute of Materials Science and Nanotechnology, National Nanotechnology Research Center (UNAM), Bilkent University, Ankara, Turkey \\
        $^{6}$ Department of Electrical and Computer Engineering, Saint Louis University, St. Louis, MO 63103, USA
        \\
		$^7$Department of Applied Physics, Yale University, New Haven, Connecticut 06520, USA \\
		$^*$hui.cao@yale.edu
	}
	
	
	\begin{abstract}
		
		Time-reversal symmetry allows waves to retrace their paths through complex media and refocus at their origin. However, incomplete capture and reversal of scattered waves often limit pulse recovery. We address this challenge for spatially extended sources by introducing a generalized time-reversal framework, which identifies the optimal source pattern as the one that maximizes the energy of the waves that are captured and reversed. \textcolor{black}{We experimentally verify it in a two-dimensional diffusive waveguide, showing the peak transmitted power at the recovery time is doubled over standard time-reversal. Internal spatiotemporal dynamics, reconstructed from frequency-resolved measurements}, reveal a ``loading and firing" process in which energy accumulates within the medium and then is released abruptly. Time-reversing this burst enables deep energy delivery into the scattering medium. Generalized time-reversal opens new possibilities for short-pulse control in scattering media, with applications ranging from optogenetics to random laser amplification.
		
	\end{abstract}
	
	\maketitle
	
	\section{Introduction}
	
\begin{figure*} 
		\includegraphics[width=\textwidth]{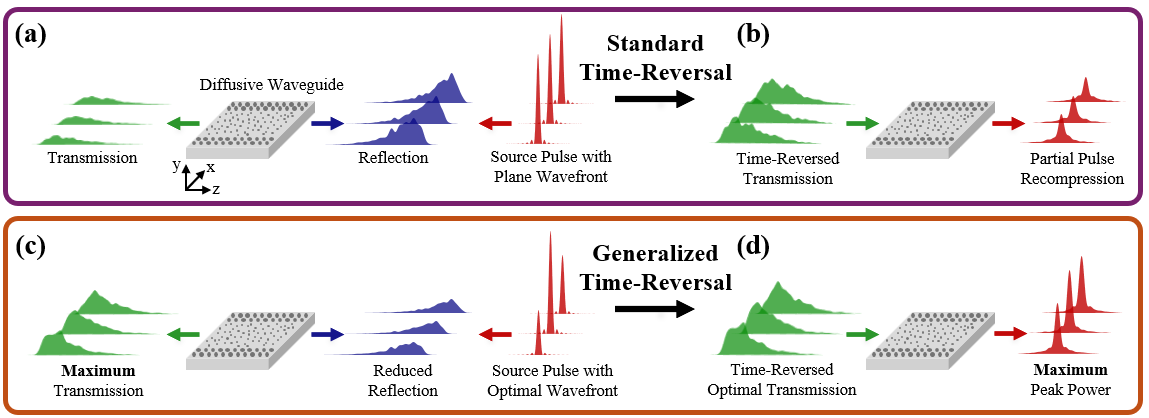}
		\caption{\label{Fig1} \textcolor{black}{\textbf {Firing a pulse through a diffusive system.} (a) A short pulse with plane-wavefront enters a diffusive waveguide from the right (red). Most incident waves are reflected (blue), and some is absorbed (not shown). The transmitted pulse (green) is temporally stretched and distorted by multiple scattering. (b) Standard time-reversal of only the transmitted waves, after renormalization to unit flux, partially recovers the pulse but with a broad background and a 10-times lower peak power than the original. (c) In generalized time-reversal, the source wavefront is first optimized to maximize the energy of transmitted waves through the waveguide by minimizing reflection and absorption. (d) Time-reversing the optimal transmitted field produces a spatiotemporal input pulse on the left that maximizes the transmitted peak power on the right, doubling the value obtained with standard time-reversal. Reflected waves are not shown in (b, d). Pulse profiles and enhancement values are calculated from numerical simulations of a 2D diffusive waveguide identical to that used in the experiment, including absorption, and are not plotted to scale.}}
	\end{figure*}

	Time-reversal (TR) symmetry is a fundamental principle of physics stating that a physical system, if reversed in time, will return to its original state. TR has been widely used to control wave propagation in complex systems such as chaotic cavities~\cite{Fink1992, kuperman1998phase, Fink2000, Lerosey2004, Sutin2006, lerosey2007focusing, Przadka2012, dezfooliyan2015spatiotemporal, delHougne2016, lerosey2022wavefront} and scattering media~\cite{Derode1998, Derode1995, Derode2002, yaqoob2008optical, Lemoult2009, McCabe2011, Aulbach2011, xu2011time, Aulbach2012, Katz2011, McCabe2011, Small2012, morales2015delivery}. Experimentally, complete TR is difficult to realize, as typically only a fraction of the outgoing waves are time-reversed. Figure~\ref{Fig1} shows an example of incomplete TR. A short laser pulse with a plane wavefront is launched from the right into an optical waveguide containing randomly distributed scatterers. After multiple scattering, most of the light is reflected, some is absorbed, and the remainder is transmitted. The transmitted pulse is stretched and distorted in time, and the pulse shape varies spatially (Fig.~\ref{Fig1}(a)). Even if the transmitted field on the left is time-reversed at every spatial location and renormalized to unit flux, sending it back to the waveguide does not fully recover the peak power of the original pulse (Fig.~\ref{Fig1}(b)), because the reflected and absorbed waves are not time-reversed.
	
	\textcolor{black}{Is it possible to outperform standard TR (sTR) while still reversing only the transmitted waves? The answer is yes. As we will show, maximum refocused power is achieved by optimizing the source pattern on the right (the spatial wavefront of the original pulse) to maximize the transmitted energy on the left (Fig.~\ref{Fig1}(c)). By minimizing the amount of energy being reflected and absorbed, the effect of incomplete TR is reduced as much as possible.} This concept is generalizable to any linear reciprocal system: the optimal source pattern for incomplete TR couples most energy to the spatial channels in which waves will be time-reversed. We refer to this principle as {\it generalized time-reversal} (gTR).
    
    \textcolor{black}{Experimentally, it is difficult to implement gTR from right to left if the wavefront shaping apparatus is available only at the input (left) side. Instead, we exploit reciprocity and measure the time-resolved spatiospectral transmission matrix from left to right~\cite{Andreoli2015, Mounaix2016, Mounaix2020TimeRO, Ferise2023}. Its dominant singular vector defines the spatiospectrally-modulated input field on the left that realizes the gTR objective of maximizing the transmitted power at a chosen time on the right, as illustrated in Fig.~\ref{Fig1}(d).}
	
	We reveal how this gTR process generates an intense pulse in transmission by reconstructing the spatiotemporal dynamics inside a two-dimensional (2D) disordered waveguide from frequency-resolved measurements. Spatiospectral shaping of a broadband input first loads energy in the first half of the diffusive waveguide, and then fires it through the second half. The resulting burst has the shortest possible duration, determined by the input field's spectral bandwidth. This process bears similarity to cavity quality ($Q$) factor switching in pulsed lasers, namely, altering the $Q$ factor for energy storage and subsequent release. In our diffusive system, however, the release of stored energy occurs not by reconfiguring the scattering structure or through any nonlinear effects, but by the manipulation of the interference of scattered waves in space and time. 

    Our theoretical model shows that the internal dynamics underlying pulse loading-and-firing are governed by long-range correlations of multiply scattered waves. The model predicts not only the maximum achievable enhancement of the transmitted peak power, but also the full internal spatiotemporal dynamics. Through direct comparison with spatial-only, spectral-only, and sequential shaping schemes, we show that spatiospectral shaping provides a level of control unattainable with other wavefront-shaping approaches.
    
    Finally, we demonstrate experimentally that time-reversing the pulse loading-and-firing process leads to deep pulse injection into a diffusive system. While our experiments are conducted with light, the framework developed here is broadly applicable to other types of waves, including microwaves, acoustics, water waves, and mesoscopic electrons.
    
	\textcolor{black}{\section{Time-resolved spatiospectral transmission matrix}}
	
	To maximize total transmitted peak power in all spatial channels, we introduce the time-resolved spatiospectral transmission matrix that maps the incident spatiospectral profile on the left to the transmitted field pattern at a delay time $t$ on the right (Fig.~\ref{Fig1}(d)). The Fourier transform of the spatiospectrally modulated field $\Psi_{\rm in}(x_i,\nu)$ gives the spatiotemporal pulse $\Psi_{\rm in}(x_i,t)$ incident on a 2D waveguide, where $x_i$ denotes the transverse coordinate at the waveguide input. The transmitted field profile at time $t$ is
	\begin{equation}
		\Psi_{\rm out}(x_o, t) = \int d x_i \int d \nu \, \mathfrak{T}(x_o, t; x_i, \nu) \, {\Psi}_{\rm in}(x_i, \nu),
        \label{EqtSSTMInt}
	\end{equation}
	where $\mathfrak{T}(x_o, t; x_i, \nu) = e^{-i \, 2 \pi \nu t} \, T(x_o, x_i; \nu)$. The transmission coefficient $T(x_o, x_i; \nu)$ relates the incident field at $x_i$ on the left to the transmitted field at $x_o$ on the right for frequency $\nu$~\cite{Vellekoop2008, Popoff2010}.
	
	$\mathfrak{T}(x_o, t; x_i, \nu)$ can be expressed in matrix form, which we term the time-resolved spatiospectral transmission matrix (tSSTM). The total number of spatial channels $N_x$ is given by the number of guided modes in the waveguide. The number of spectral channels is $N_\nu = 1+\Delta \nu / \delta \nu$, where $\Delta \nu$ is the spectral bandwidth of an input pulse, and $\delta \nu$ is the spectral correlation width of the transmitted field through the waveguide (see Methods and Supplementary Materials (SM) Sec.~\ref{Sec:linewidth}~\cite{methods}). The matrix $\mathfrak{T}(t)$ has dimension $N_x \times N_x N_\nu$. The spatiospectrally modulated input  $\vert\Psi_{\rm in}\rangle$ is described by a vector with $N_x N_\nu$ elements, and the transmitted field $\vert\Psi_{\rm out}\rangle$ at $t$ is a vector of $N_x$ elements.
	
	The transmitted power from left to right ($L\to R$) at time $t$ for an incident spatiospectral wavefront $\vert\Psi_{\rm in}\rangle$ is given by $\langle \Psi_{\rm in} | \mathfrak{T}^\dagger(t) \mathfrak{T}(t) | \Psi_{\rm in} \rangle$. Therefore, the largest eigenvalue of $\mathfrak{T}^\dagger(t) \mathfrak{T}(t)$ maximizes the transmitted power at $t$, and the corresponding eigenvector is the optimal input field.
    
    \textcolor{black}{\section{Experiments}}
    
	We conduct experiments in a 2D silicon waveguide with randomly distributed air holes (see Methods and SM Sec.~\ref{Sec:Exp}~\cite{methods}). Detection of out-of-plane scattered light allows us to monitor light propagation in the 2D waveguide from the third dimension~\cite{Sarma2016}. The disordered region has length $L = 50$ \textmu m and width $W = 15$ \textmu m. Both dimensions are much larger than the transport mean free path $\ell_t = 3.3$ \textmu m at optical wavelength $\lambda \approx 1550$ nm ~\cite{Sarma2015_2}; thus, light transport is diffusive. \textcolor{black}{The number of spatial channels is $N_x \approx 54$.} While material absorption of light is negligible, out-of-plane scattering by the air holes causes weak dissipation, as described by the diffusive dissipation length $\xi_a = 28$ \textmu m~\cite{Yamilov2014}. 
    
\begin{figure*}
		\includegraphics[width=.75\textwidth]{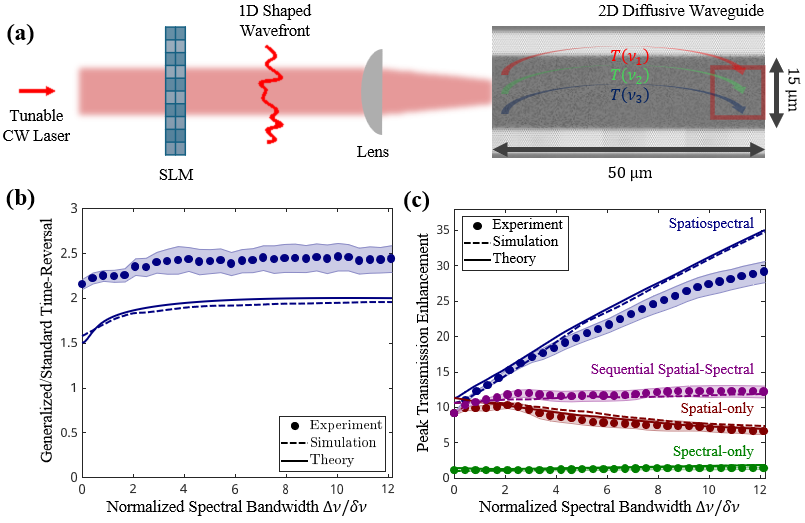}
		\caption{\label{Fig2} \textcolor{black}{\textbf {Experimental reconstruction and peak-transmission enhancement.} (a) A tunable continuous-wave (CW) laser is spatially shaped by an SLM and injected into a 2D diffusive waveguide in silicon-on-insulator (SOI). Frequency-resolved transmission matrices $T(\nu_j)$ are measured by scanning the laser frequency to construct the tSSTM mapping the left input fields to fields in a 10-by-10 \textmu m$^2$ area at the right end (red). A scanning electron microscope image of the fabricated waveguide is shown. (b) Enhancement of generalized time-reversal over standard time-reversal, given by the ratio of their peak transmitted powers, as a function of the spectral bandwidth $\Delta\nu$ normalized to the spectral channel width $\delta\nu$. (c) Enhancement of peak transmitted power relative to an unmodulated transform-limited pulse for spatiospectral (blue), sequential spatial-spectral (purple), spatial-only (red), and spectral-only (green) shaping. Spatiospectral control outperforms all other methods. In (b, c), experimental data are shown by circles, numerical simulations by dashed lines, and theoretical predictions by solid lines. The shaded regions represent one standard deviation about the mean for nine repeated measurements. }}
	\end{figure*}

	\textcolor{black}{Instead of spatiospectral shaping of an optical pulse, we scan the input frequency of a continuous wave to acquire the frequency-resolved response and then reconstruct spatiotemporal dynamics with the Fourier transform (Fig.~\ref{Fig2}(a)). Using a frequency-tunable laser, we measure the frequency-resolved transmission matrices~\cite{Bender2020}, and construct the tSSTM $\mathfrak{T}(t)$ (see Methods and SM Sec.~\ref{Sec:Exp}~\cite{methods}). We compute the largest eigenvalue of $\mathfrak{T}^\dagger(t)\mathfrak{T}(t)$ and compare it with the transmitted peak power for a transform-limited pulse with an unmodulated wavefront. Their ratio defines the enhancement factor $\eta$, which reaches 35 for $N_x$ = 54 and $N_\nu$ = 17, corresponding to a total of $N_x \, N_\nu$ = 918 spatiospectral channels. Figure~\ref{Fig2}(c) shows that $\eta$ increases with the spectral bandwidth $\Delta \nu$, owing to the constructive interference of an increasing number of spectral channels at the target time. }

	\section{Generalized Time-Reversal}
	
	The $L\to R$ tSSTM provides the incident wavefront that maximizes the peak power of the total transmitted field at a target time on the right, thereby achieving temporal focusing across all output channels. This approach is distinct from conventional spatiotemporal focusing: launching a transform-limited pulse from a point source on the right and time-reversing the transmitted fields on the left. This yields a wavefront that refocuses at the original source location and maximizes the peak power there. However, this method does not directly generalize to multiple foci, because
    the relative amplitudes and phases of the initial sources that would maximize the total peak power via TR are not known {\it a priori}. To identify the optimal source pattern for time-reversal in this general setting, we discover a general relation between the tSSTM and the broadband transmission matrix (BTM)~\cite{Hsu2015}.

	We prove that the tSSTM from $L\to R$ of a disordered waveguide is related to the BTM from right to left ($R\to L$). First, we note that $\mathfrak{T}^\dagger(t) \mathfrak{T}(t)$ and $\mathfrak{T}(t) \mathfrak{T}^\dagger(t)$ share the same non-zero eigenvalues. We then express the latter as
	\begin{equation}
		\mathfrak{T}(t) \mathfrak{T}^\dagger(t) = \frac{1}{N_\nu}\sum_{j=1}^{N_\nu} T_j T_j^\dagger =\!\! \left[\frac{1}{N_\nu}\sum_{j=1}^{N_\nu} \!\left(T_j^T\right)^\dagger \!T_j^T\right]^* \!\!\!\! = \!(\mathcal{A}')^*,
		\label{SSTM_BTM_equiv}
	\end{equation}
	where $T_j$ denotes the $L\to R$ transmission matrix for the $j$-th spectral channel, $j = 1, 2, \ldots N_\nu$. By reciprocity, its transpose $T_j^T$ gives the $R\to L$ transmission matrix~\cite{Beenakker1997}. The $R\to L$ BTM is defined as $\mathcal{A}' \equiv (1/N_\nu) \sum_{j=1}^{N_\nu} \left(T_j^T\right)^\dagger T_j^T$, where the prime superscript denotes a mapping from $R\to L$. Its largest eigenvalue gives the maximum time-integrated transmission over all spatial channels under spatial wavefront shaping of an input pulse with $N_\nu$ spectral channels~\cite{Hsu2015, McIntosh2024}. 
	
	Equation~\eqref{SSTM_BTM_equiv} ensures that the maximum eigenvalue of $\mathfrak{T}^\dagger(t) \mathfrak{T}(t)$ is equal to that of $\mathcal{A}'$, independent of the target time $t$. This implies that the highest transmitted peak power achieved by spatiospectral shaping in the $L\to R$ direction is proportional to the maximum time-integrated transmission obtained by spatial-only shaping in the $R\to L$ direction. The corresponding eigenvector of $\mathcal{A}'$ defines the optimal source pattern (spatial wavefront) for gTR, which maximizes the total energy of the transmitted fields that will be time-reversed on the left. After launching a transform-limited pulse with this spatial wavefront from the right, the transmitted field at the left is proportional to the phase conjugate of the eigenvector of $\mathfrak{T}^\dagger(t) \mathfrak{T}(t)$ with the largest eigenvalue. Time-reversing the transmitted pulse at the left then minimizes the impact of not reversing the reflected and absorbed waves, thereby maximizing the peak transmitted power at the right. This result generalizes TR from spatiotemporal focusing in a single channel to global temporal focusing in all spatial channels. Additional details are provided in SM Sec.~\ref{Sec:GTR}~\cite{methods}.

    \textcolor{black}{Using the experimentally constructed tSSTM, we quantify the enhancement of gTR over sTR. For sTR of a source pulse with an unoptimized spatial wavefront $\vert\psi_\text{S}\rangle$ on the right, the transmitted waves at the left are time-reversed to produce a spatiospectral input pulse given by $\mathfrak{T}^\dagger(t)\vert\psi_\text{S}\rangle^*$. For $N_x$ = 54 and $N_\nu$ = 17, sTR produces an intensity enhancement of about 17 over that of a transform limited pulse. The global focusing efficiency, defined as the ratio of power at the target time to the total transmitted power in the $L\to R$ direction, is 0.47. These results are obtained after averaging over 100 random spatial wavefronts $\vert\psi_\text{S}\rangle$. By additionally optimizing $\vert\psi_\text{S}\rangle$ to maximize the energy of transmitted fields before time-reversal, gTR achieves an enhancement of about 35, and a focusing efficiency of 0.69, representing an approximately twofold increase in peak power and a nearly $50\%$ increase in focusing efficiency over sTR. In Fig. 2b, we plot the enhancement in peak power of gTR over sTR as a function of bandwidth. In the monochromatic limit ($\Delta\nu = 0$), the average transmitted power obtained by sTR reduces to the well-known value $(2/3) T_\text{max}$, where $T_\text{max}$ is the maximum transmission eigenvalue for a monochromatic wave~\cite{Vellekoop2008, Yu2013}. Therefore, the ratio between $T_\text{max}$ and the average transmitted power is $3/2$. As $\Delta\nu$ increases, the theoretical gTR over sTR enhancement increases from $1.5$ to $2$, in agreement with numerical simulations (more details in SM Sec.~\ref{Sec:Sim}). These results are discussed in Secs.~IV and V and derived in SM Secs.~\ref{Sec:GTR} and~\ref{Sec:MP}. The experimentally reconstructed enhancement is slightly larger, possibly due to the lower $R\to L$ transmitted power and hence greater noise sensitivity of sTR than of gTR~\cite{Yilmaz2019}. This result further illustrates the advantage of gTR over sTR.}

	\textcolor{black}{\section{Theoretical model}}
		
	We develop a theoretical model to predict peak power enhancement by spatiospectral shaping. Using Eq.~\eqref{SSTM_BTM_equiv}, we derive the maximum transmitted peak power from the largest eigenvalue of the backward BTM calculated with filtered random matrix theory~\cite{Goetschy2013, Hsu2015, McIntosh2024}. The value is normalized by the peak transmission of a transform-limited pulse with a random spatial wavefront (see Methods and SM Sec.~\ref{Sec:Unmodulated}~\cite{methods}) to obtain the enhancement over unmodulated pulses. 
    
    The enhancement factor $\eta$ is determined by the effective number of independent spatiospectral channels $N_\text{eff}$, defined as $1/N_\text{eff}=\bar{C}_1/N_x + \bar{C}_2$, where $\bar{C}_1$ and $\bar{C}_2$ are the frequency-integrated short- and long-range contributions to the intensity correlation (see Methods). For spatiotemporal focusing into a single spatial channel, the long-range correlation $\bar{C}_2$ is negligible, and $N_{\text{eff}}=N_x/\bar{C}_1=N_x  N_\nu$ directly gives the focusing enhancement~\cite{Lemoult2009}. By contrast, for global temporal focusing across all spatial channels, long-range contributions dominate and $N_\text{eff} \approx 1/\bar{C}_2
	\sim N_x\sqrt{ N_\nu} \, (\ell_t/L)$. \textcolor{black}{The peak power enhancement then scales as 
	\begin{equation}
		\eta \sim N_\nu\left( 1+ \sqrt{N_x\bar{C}_2} \right)^2 \sim \frac{L}{\ell_t} \sqrt{N_{\nu}} \sim \frac{L^2}{\sqrt{\ell_t^3 \, c/\Delta\nu}},
		\label{enhScaling}
	\end{equation}
    and is therefore independent of the number of controlled spatial channels $N_x$ for complete spatial control.  Naively the  enhancement is expected to increase linearly with $N_\nu$ upon adding spectral control to spatial control of monochromatic light~\cite{Bender2022_2}, but Eq.~\eqref{enhScaling} yields  $\sqrt{N_\nu}$ scaling. A complete derivation is provided in SM Sec.~\ref{Sec:MP}~\cite{methods}. The predicted dependence of $\eta$ on $N_\nu$ agrees well with the experimental and numerical  results shown in Fig.~\ref{Fig2}(c) (see Methods and SM Sec.~\ref{Sec:Sim}~\cite{methods} for numerical details). 
    Our theory also reveals a qualitative distinction between peak-transmission enhancement and broadband transmission enhancement~\cite{McIntosh2024}. While the latter is essentially independent of sample thickness (waveguide length $L$), the former increases quadratically with $L$, indicating that spatiospectral control becomes increasingly advantageous in thicker diffusive systems.}
    
    \textcolor{black}{Furthermore, we demonstrate in Fig.~\ref{Fig2}(c) that spatiospectral shaping, which assigns independent spatial wavefronts to individual spectral channels, outperforms spatial-only (red), spectral-only (green), and sequential spatial–spectral shaping (purple) of input pulses (see Methods). For spatial-only shaping, all spectral channels share the same wavefront and form a transform-limited pulse. Owing to the absence of spectral control, the maximum peak-transmission enhancement decreases with increasing bandwidth and scales as $1/\sqrt{N_\nu}$, corresponding to a factor $1/N_\nu$ relative to spatiospectral control.
    For spectral-only shaping, all spatial channels share the same optimized spectrum and temporal profile. The enhancement arises primarily from compensation of spectral dispersion induced by multiple scattering and, in the broadband limit, scales as $\sqrt{N_\nu}/N_x$, reflecting the absence of spatial control. The enhancement factors for both spatial-only and spectral-only shaping are derived in SM Sec.~\ref{Sec:MP}~\cite{methods}.
    To illustrate that spatiospectral shaping is not separable into spatial and spectral shaping, we consider successive optimization in which a single spatial wavefront is first optimized across all spectral channels, followed by optimization of a single spectrum across all spatial channels (see Methods). The sequential spatial–spectral shaping results in an enhancement nearly independent of bandwidth and strictly below that of simultaneous spatiospectral shaping. This comparison highlights that joint spatial and spectral optimization exploits spatiotemporal correlations that cannot be accessed through separable optimization strategies.}
    \\
	\section{Pulse loading and firing}
	
	\begin{figure*} 
		\includegraphics[width=.85\textwidth]{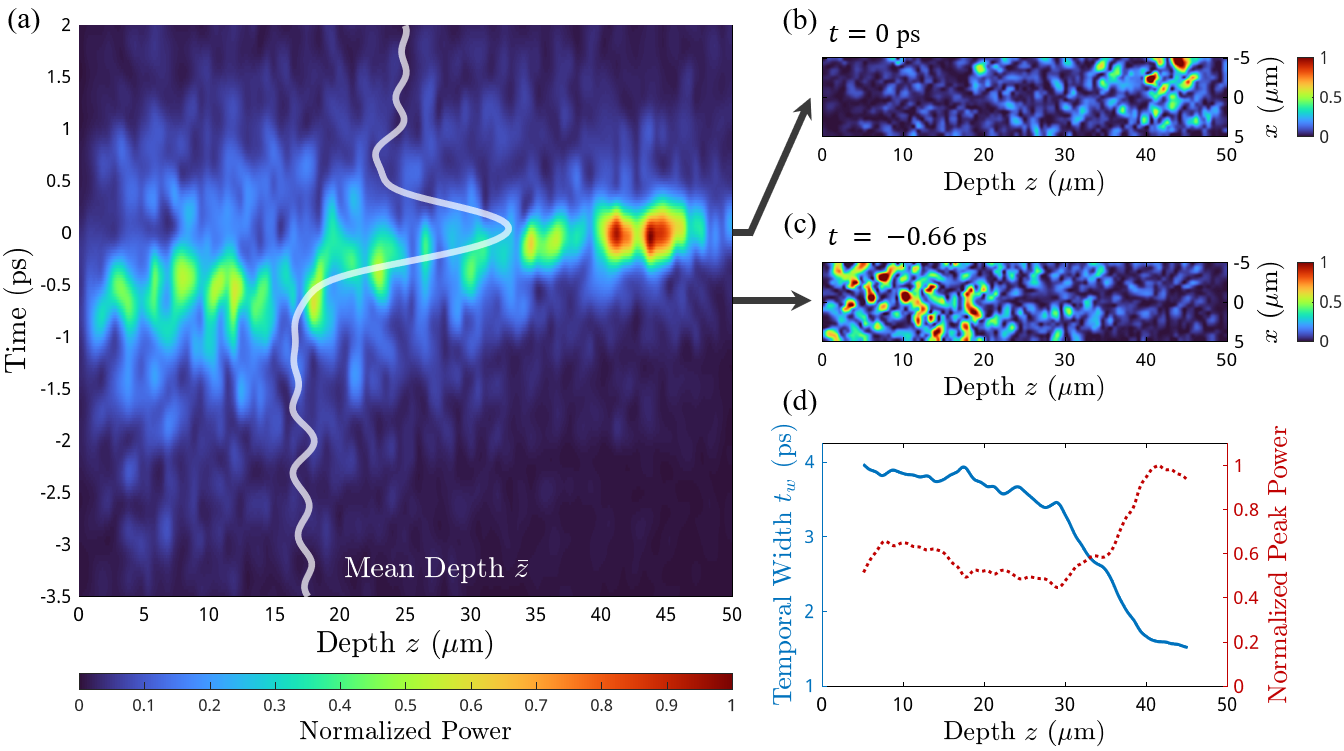}
		\caption{\label{Fig3} \textbf {Experimental demonstration of pulse loading and firing.} (a) Spatiotemporal power distribution $P(z,t)$ of the dominant spatiospectral transmission eigenchannel in a diffusive waveguide. Optical power is accumulated in the front half of the waveguide during the loading stage $t \leq -0.66$ ps, then quickly moves to the waveguide exit in the firing stage to generate a pulse in transmission at $t = 0$. The color scale is normalized to the maximum value at $t=0$. The spectral bandwidth of the input pulse is $\Delta\nu$ = 2.45 THz centered at frequency 191.56 THz. Mean depth $\bar{z}(t)$ of internal power distribution is plotted by the white line, revealing the transition from loading to firing at $t \simeq -0.66$ ps. (b) Intensity distribution $I(x,z,t)$ at $t = 0$ showing that the waveguide exit is lit up. (c) Intensity distribution at $t = -0.66$ ps showing that energy is accumulated in the front half. For better visualization, images in (b, c) are slightly saturated by normalizing intensity to 2/3 of their maxima. (d) Temporal pulse width $t_w(z)$ (blue) and the peak power, normalized to the maximum (red dotted) as a function of depth $z$, indicating that loading dominates at $z \lesssim 30$ \textmu m and firing at $z\gtrsim 30$ \textmu m. The peak power at the waveguide exit is enhanced 35-fold over an unmodulated transform-limited pulse.}
	\end{figure*}
	
	To reveal how spatiospectral shaping transforms a diffusive waveguide into a pulse compressor, we reconstruct the internal spatiotemporal dynamics from frequency-resolved measurements of the field distribution everywhere inside the 2D waveguide (SM Secs.~\ref{Sec:Exp} and~\ref{Sec:GTRvsTR}~\cite{methods}). Supplementary Video 1 reveals pulse loading and firing through the diffusive sample to maximize the peak transmitted power at a target time $t = 0$ ps. For the dominant eigenvector of $\mathfrak{T}(t)^\dagger \mathfrak{T}(t)$, we integrate intensity $I(x,z,t)$ in the waveguide cross section and plot the power $P(z,t) = \int_0^W I(x,z,t) \, dx$ in Fig~\ref{Fig3}(a) as a function of depth $z$ and time $t$. As the pulse enters the disordered waveguide, \textcolor{black}{energy builds up inside through destructive interference between waves leaking out of the waveguide and reflected waves from the front surface}. Near the end of this loading stage, $t \simeq -0.66$ ps, energy is accumulated in the front portion of the waveguide, shown in Fig.~\ref{Fig3}(c). Then, the stored light moves quickly toward the exit, creating a giant pulse in transmission, \textcolor{black}{via constructive interference of forward-scattered waves}. Figure~\ref{Fig3}(b) shows that the region close to the waveguide exit is lit up at the target time $t=0$. 
	
	To quantify the transition from loading to firing, we calculate the mean axial position $\bar{z}(t) = \int P(z,t) \, z \, dz /\int P(z,t) \, dz$ as a function of time. $\bar{z}(t)$ is overlaid on the spatiotemporal profile $P(z,t)$ in Fig.~\ref{Fig3}(a) (white line). At early times, $t < -0.66$ ps, $\bar{z}(t)$ is invariant, as energy is built up inside the sample. Once the pulse firing process begins, $\bar{z}(t)$ increases rapidly. It then falls back as the pulse exits the waveguide. The transition from loading to firing can also be observed from the temporal pulse width $t_w(z)$ (see Methods) as a function of depth $z$ in Fig.~\ref{Fig3}(d). For $z\lesssim 30$ \textmu m, $t_w(z)$ remains relatively large and invariant with $z$, indicating that the loading phase dominates in the front half of the waveguide. For $z\gtrsim 30$ \textmu m, $t_w(z)$ drops rapidly, reflecting that the pulse is temporally compressed as it moves toward the exit. The red curve in Fig.~\ref{Fig3}(d) shows that the peak power remains low and nearly constant for $z\lesssim 30$ \textmu m, but grows quickly for $z\gtrsim 30$ \textmu m. 
	
	The loading-and-firing process also occurs in spatiotemporal focusing to a single output channel, as confirmed by our numerical simulation, but the effect is substantially weaker than global temporal focusing. We analytically reproduce loading and firing by calculating the internal profile $P(z,t)$ for spatiotemporal focusing at $z=L$ using sTR,
    \begin{equation}
    \label{EqDecompoIntensity}
        P(z,t) = P_f(z,t)+ P_s(z,t)+ P_l(z,t).
    \end{equation}
	Explicit analytical expressions for these terms and associated spatiotemporal profiles are presented in SM Sec.~\ref{Sec:GTRvsTR}~\cite{methods}. $P_f(z,t)$ contributes mainly to the focus at $z=L$, and $P_s(z,t)$ is concentrated near the front surface $z=0$. $P_l(z,t)$ results from long-range intensity correlations between $z$ and $L$, and dictates the internal power distribution away from the front and back surfaces. While the spatiotemporal focusing enhancement is dominated almost entirely by short-range correlations $C_1$, the internal spatiotemporal dynamics is actually dictated by long-range correlations $C_2$. Similarly, for global temporal focusing, pulse loading and firing are also governed by long-range intensity correlations of diffuse waves. Our analysis further reveals that light dissipation, as in our sample, shortens the loading process and weakens the fired pulse (SM Sec.~\ref{Sec:Abs}~\cite{methods}).
	
	\section{Deep Pulse Injection}
	
	\begin{figure*} 
		\includegraphics[width=\textwidth]{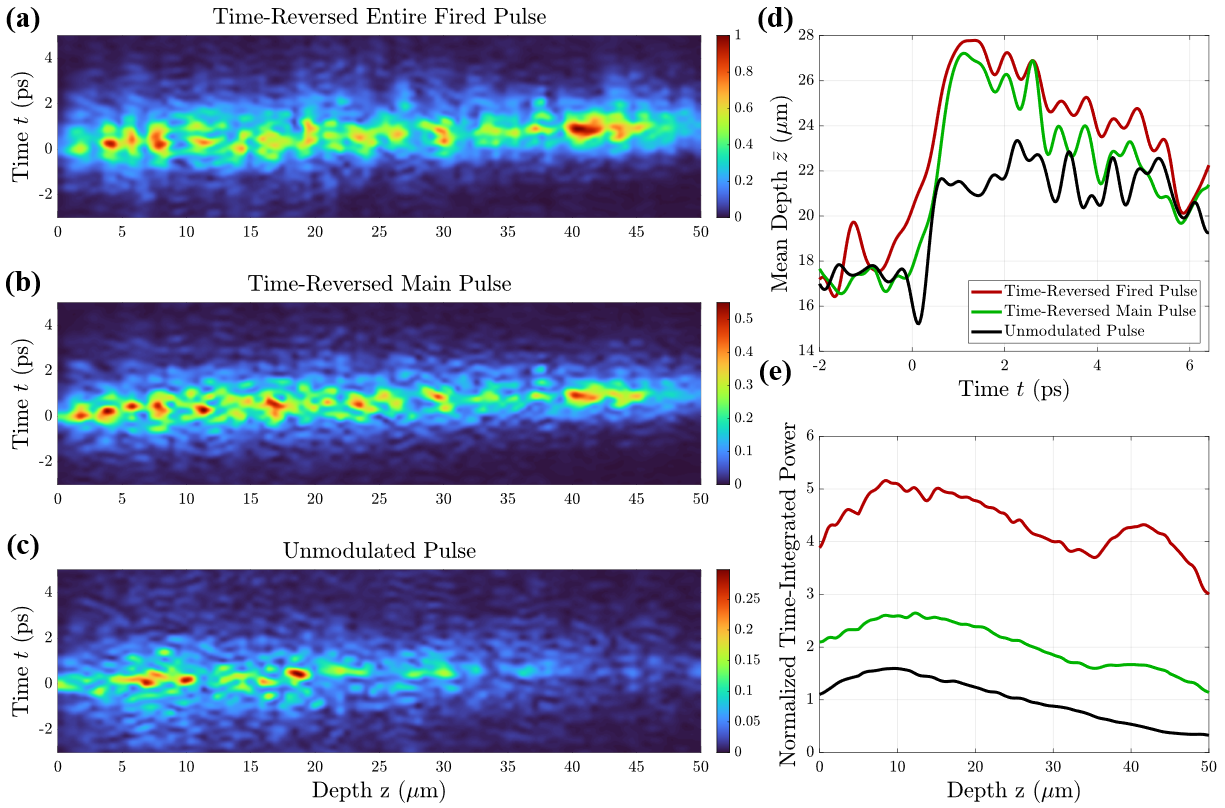}
		\caption{\label{Fig4} \textcolor{black}{\textbf {Pulse injection via time-reversed loading and firing.} (a) Experimental reconstruction of the spatiotemporal power distribution $P(z,t)$ inside the waveguide after time-reversing the fired pulse at the left side. The spatiotemporally-modulated pulse is injected from the left at $t=0$ deep into the waveguide. (b) Time-reversing only the spatial wavefront at $t = 0$ injects a transform-limited pulse into the sample, corresponding to time-reversing only the main pulse.  (c) A transform-limited pulse with unmodulated plane-wavefront and identical bandwidth remains closer to the input side. (d) Mean depth $\bar{z}$ of the internal power distribution as a function of time when time-reversing the entire fired pulse (red) in (a), only the main pulse (green) in (b), and spatially-unmodulated pulse (black) in (c). (e) Depth profile of time-integrated internal power normalized to that of the unmodulated pulse averaged over depth for (a--c). Time-reversal of entire fired pulse (a) delivers 4-5 times more energy than spatially-unmodulated pulse (c). A moving average over depth is applied using a window of width $0.2L$. The depth is normalized by the waveguide length $L$, and the color scales in (a--c) are normalized to the maximum value in (a). The enhanced power delivery at $z>0.8L$ in (a) results from approximating the transmitted field by that in a region next to the waveguide exit.}}
	\end{figure*}
	
	\textcolor{black}{The loading-and-firing dynamics suggests the possibility of deep energy delivery into a diffusive system by time-reversing the fired pulse. Since the firing results from a coordinated release of energy stored deep inside the waveguide, its time-reversal corresponds to deep pulse injection before time-reversed loading starts. To inject the fired pulse from $L\to R$, we experimentally construct the $R\to L$ tSSTM, given by replacing frequency-resolved transmission matrices with their transpose $T(\nu)\to T^T(\nu)$ in Eq.~\eqref{EqtSSTMInt}. Its dominant singular vector defines the incident wavefront on the right that produces the strongest fired pulse at the left. The transmitted field from right to left is time-reversed and then injected back into the waveguide from the left. The resulting spatiotemporal field evolution inside the waveguide is then reconstructed from frequency-resolved measurements.} 
    
    \textcolor{black}{The internal power distribution $P(z,t)$ in Fig.~\ref{Fig4}(a) shows deep pulse injection into the diffusive waveguide, in contrast to shallow penetration of a transform-limited pulse with plane wavefront and identical bandwidth in Fig.~\ref{Fig4}(c). This is further illustrated in the mean depth $\bar{z}$ plotted in Fig.~\ref{Fig4}(d). Time-integrated internal power as a function of depth in Fig.~\ref{Fig4}e reveals that time-reversing the fired pulse enhances the injected power by a factor of 4--5 throughout the waveguide.    
    Since the transmitted pulse created by loading and firing has a broad background, its time-reversal injects light beyond the main pulse with a time-varying spatial wavefront, which can be implemented by spatiospectral shaping. To realize spatial-only shaping of a transform-limited pulse for deep injection, we find the optimal spatial wavefront by phase-conjugating the transmitted field pattern of the main pulse at $t=0$. According to gTR, this field pattern maximizes time-integrated transmission in the $L \to R$ direction. The spatiotemporal power distribution in Fig.~\ref{Fig4}(b) shows a much deeper pulse injection into the waveguide than the input pulse with an unmodulated wavefront in (c). While the internal profile of time-reversing the main pulse is similar to that of time-reversing the entire fired pulse (Fig.~\ref{Fig4}(d)), the time-integrated injected power is reduced by half as shown in Fig.~\ref{Fig4}e. This difference results from additional energy delivery through the time-reversed background beyond the main pulse. Supplementary Video 2 shows the experimental reconstruction of deep pulse injection and videos 3 and 4 show our numerical simulations of a larger waveguide without dissipation which provides clearer visualization of the time-reversed firing followed by time-reversed loading, as well as the larger difference between time-reversal of the entire fired pulse and reversal only of the spatial wavefront at the target time for a transform-limited pulse (see SM Sec.~\ref{Sec:DPI}).}
	
	\section{Discussion}
	
	Generalized time-reversal revises the conventional paradigm of time-reversal as a passive wave-refocusing protocol. By optimizing the incident wavefront prior to time-reversal, it identifies excitation conditions that maximize the subsequent spatiotemporal refocusing efficiency in complex media. Since it is independent of dimensionality, generalized time-reversal applies equally to three-dimensional (3D) diffusive systems. Despite long-range correlations being much weaker than in 2D, they still dominate pulse loading and firing, enabling deep injection of pulses in 3D. More generally, the intimate connection between long-range correlations and spatiotemporal dynamics suggests that boundary shaping or geometric confinement can provide additional degrees of control over pulse propagation by engineering these correlations~\cite{koirala2017inverse}.

	We emphasize that the principle of generalized time-reversal, demonstrated here for pulse transmission through disordered media, is not limited to this specific problem. Rather, it provides a general framework for optimal spatiotemporal wave control in any linear reciprocal system with specified input and output ports. Consequently, it applies naturally to wave control in reflection~\cite{Choi2013, Jeong2018, lambert2020reflection, balondrade2024multi}, energy deposition~\cite{Durand2019, Bender2022_2, McIntosh2024}, remission~\cite{Bender2022}, and multi-target control~\cite{Shaughnessy2024}. The same principle also extends to systems with partial order~\cite{Uppu2021}, spatial correlations or symmetry~\cite{aubry2020experimental, Davy2021, saini2024mirror}, and engineered transmission matrices~\cite{Dinsdale2021, Horodynski2022}. Beyond linear systems, generalized time-reversal may also offer new opportunities to control nonlinear wave dynamics~\cite{fisher1983optical, chabchoub2014time, ducrozet2020experimental}, where optimized energy loading and release could enable phenomena such as giant pulse generation in random laser amplifiers.

	
	\section*{Data Availability Statement}
	
	The data that support the findings of this study are available from the corresponding author upon reasonable request.
	
	\section*{Acknowledgments}
	
	We acknowledge stimulating discussions with the late Chia-Wei Hsu. This work is supported partly by the grants DMR-1905465 and DMR-1905442 from the US National Science Foundation (NSF), by the grant 2024-337821 from the Chan Zuckerberg Initiative (CZI), and by the program Investissements d’Avenir from the French Government.  
	
	\section*{Author Contributions}
	
	R.M. performed the experiments and analyzed the data. N.B. fabricated the samples. H.Y. contributed to the experimental analysis. A.G. and R.M. developed the analytic theory. R.M. and A.Y. performed the numerical simulations. H.C. initiated the project and supervised the research. All authors contributed to the interpretation of the results. R.M. and A.G. prepared the manuscript, H.C. edited it, and all authors provided feedback.
	
	\section*{Declaration of Interests}
	
	The authors declare no competing interests.
	
	\section*{Methods}
	
	\textbf {Experiments}. Our experiments are conducted on 2D waveguides fabricated in silicon-on-insulator wafers using electron beam lithography and reactive ion etching. Light scattering inside the waveguide is induced by randomly distributed air holes with 100 nm diameter and a filling fraction of $5.5\%$. The waveguide sidewalls are composed of highly reflecting photonic crystal layers to avoid light leakage. A detailed schematic of our experimental setup is provided in SM Sec.~\ref{Sec:Exp}. A monochromatic laser beam with a tunable wavelength around 1550 nm (Keysight 81960A) is modulated by a phase-only spatial light modulator (Hamamatsu LCoS X10468) and then injected into one waveguide via the edge of the wafer. Vertically scattered light from the disordered region is interfered with a reference beam, and the interference pattern is recorded by a CCD camera (Allied Vision Goldeye G-032 Cool). The complex field distribution inside the sample is retrieved from the interference patterns in four successive measurements with SLM phase shifts of $0$, $\pi/2$, $\pi$, and $3\pi/2$. \textcolor{black}{The transmission matrix is approximated by mapping the input fields to the field distribution in a $10\times10$~\textmu m$^2$ region at the output end of the waveguide (Fig. 2a).} We measure 41 frequency-resolved transmission matrices over a frequency range of $\Delta \nu = 2.45$ THz~\cite{Bender2020}, and then construct the tSSTM. From the frequency-resolved mapping of incident fields to the internal field distribution everywhere inside the waveguide, we reconstruct pulse propagation in time for any input wavefront. In Fig.~\ref{Fig3}(d), the temporal pulse width is defined as the participation number, $t_w(z) = \left[ \int P(z,t) \, dt \right]^2/\int \left[ P(z,t) \right]^2  \, dt$. Additional details on the experimental setup and the tSSTM measurement are given in SM Sec.~\ref{Sec:Exp}~\cite{methods}.
	
	\textbf {Numerical Simulations}. We simulate pulse propagation in 2D disordered waveguides using KWANT, an open-source Python package for scalar wave transport simulations with a tight-binding model~\cite{Groth2014}. In Fig.~\ref{Fig2}, the simulated waveguide parameters are identical to the experimental values. To compute the peak power enhancements, we average over 100 realizations of disorder in the waveguide. 
    Spatiotemporal profiles are generated for single realizations of disorder. To resolve the internal field distribution, a matrix mapping the incident fields to the field everywhere inside the waveguide is computed at each frequency~\cite{Bender2022_2, McIntosh2024, McIntosh2025}. The spectral Fourier transform then recovers the internal field distribution as a function of time. We conduct simulations with varying waveguide dimensions, scattering strength, and dissipation. The Thouless time $\tau_\text{Th}$ is defined as $\tau_\text{Th} = L^2/\pi^2D$, where $D$ is the diffusion constant, and is calculated from the decay in the arrival time distribution which scales as $e^{-t/\tau_\text{Th}}$. Additional details are provided in SM Sec.~\ref{Sec:Sim} and~\ref{Sec:Unmodulated}~\cite{methods}. Our numerical results show that the spatiotemporal dynamics for pulse loading and firing as well as deep injection depend on only three parameters: waveguide length $L$, Thouless time $\tau_\text{Th}$, and diffusive dissipation length $\xi_a$ (SM Sec.~\ref{Sec:GTRvsTR},~\ref{Sec:Abs}, and ~\ref{Sec:DPI}~\cite{methods}).
	
	\textbf {Spectral channel width}. $\delta \nu$ is defined by $C_1(\delta \nu/2) = (1/e) \, C_1(0)$, where $C_1(|\nu_1 - \nu_2|) =  | \overline{ T_{ji}^*(\nu_1) \, T_{ji}(\nu_2) } |^2/|\overline{ T_{ji}(\nu_1)}|^2|\overline{ T_{ji}(\nu_2)}|^2$ is the short-range spectral correlation function of the transmitted field $T_{ji}(\nu)$ at frequency $\nu$. $\delta \nu$ determines the number of uncorrelated spectral channels, $N_\nu = 1+\Delta\nu/\delta\nu$. As shown in SM Sec.~\ref{Sec:linewidth}~\cite{methods}, $1/ N_\nu = \bar{C}_1 = \iint_{\Delta\nu} d\nu_1 d\nu_2\, C_1(|\nu_1 - \nu_2|)/\Delta\nu^2 $.
	
	\textbf {Unmodulated pulse transmission}. The transmitted peak power of a transform-limited pulse with a random input wavefront and unit flux is determined by the maximal value of the arrival time distribution of transmitted light $\mathcal{F}(t)$, $P_{\text{un}}(L,t^*) = \bar{T} \mathcal{F}(t^*)$, where $\bar{T}$ is the mean transmittance. For a diffusive system, $\mathcal{F}(t)$ attains its maximum value at delay time $t^*$, which is close to the Thouless time $\tau_{\text{Th}}$. In the broadband limit of an input pulse, $\Delta \nu \gg \delta \nu$, the temporal shape of $\mathcal{F}(t)$ becomes nearly independent of $\Delta \nu$, and its amplitude scales as $1/N_\nu$. A detailed derivation is presented in SM Sec.~\ref{Sec:Unmodulated}~\cite{methods}. 
	
	\textbf {Spatiospectral shaping}. Since the spatiospectrally modulated pulse can be translated in time by a linear spectral phase ramp, the maximum eigenvalue of $\mathfrak{T}(t)^\dagger \mathfrak{T}(t)$ does not depend on the specific $t$, and gives the highest possible peak power of a transmitted pulse. Using Eq.~\ref{SSTM_BTM_equiv}, we calculate the peak transmitted power by spatiospectral shaping from the largest broadband transmission eigenvalue~\cite{Hsu2015, McIntosh2024}, 	
	$\text{max}_t[P(L,t)] = \bar{T} \left( 1+ \sqrt{N_x/N_\text{eff}} \right)^2$,
	where $\bar{T}$ is the mean transmittance for monochromatic light with random input wavefronts. $N_\text{eff}$ is determined by $\bar{C}_1$ and  $\bar{C}_2 = \iint_{\Delta\nu} d\nu_1 d\nu_2\, C_2(|\nu_1 - \nu_2|)/\Delta\nu^2$, $1/N_\text{eff} = \bar{C}_1/N_x + \bar{C}_2$. This is derived in SM Sec.~\ref{Sec:MP}~\cite{methods}. $C_{1}$ and $C_{2}$ are the short- and long-range contributions to the spectral correlation function of total transmission, $\mathcal{C}(\vert\nu_1 - \nu_2\vert)=\overline{\mathcal{T}_i(\nu_1) \mathcal{T}_i(\nu_2)}/\overline{\mathcal{T}_i(\nu_1)}\ \overline{\mathcal{T}_i(\nu_2)}-1$, where  $\mathcal{T}_i(\nu)=\sum_j \vert T_{ji}(\nu)\vert^2$ and averaging is performed over disorder realizations and spatial channels~\cite{Akkermans2007}. The peak power enhancement is defined relative to the transmitted peak power of an unmodulated transform-limited pulse, $\eta\equiv\text{max}_t[P(L,t)]/\text{max}_t[P_{\text{un}}(L,t)]$. Our numerical simulations reveal that the full-width-at-half-maximum (FWHM) of the transmitted pulse is 0.45 ps, which is approximately the inverse of its spectral bandwidth (0.41 ps).
	
	\textbf {Spatial-only shaping}. For spatial-only wavefront shaping, the incident field can be written as ${\Psi}_{\rm in}(x_i, \nu) = \psi_x(x_i) \, \psi_\nu(\nu)$. The input pulse spectrum $\psi_\nu(\nu)$ is preset, and the transmitted field is $\Psi_{\rm out}(x_o, t) = \int d x_i \, \mathfrak{T}_x(x_o, t; x_i) \, \psi_x(x_i)$, where $\mathfrak{T}_x(x_o, t; x_i) = \int d \nu \, e^{-i \, 2 \pi \, \nu t} \, T(x_o, x_i; \nu) \, \psi_\nu(\nu)$. In the matrix form, $\mathfrak{T}_x(t)$ is an $N_x \times N_x$ matrix, and $\psi_x(x_i)$ is a vector with $N_x$ elements. This recovers the time-gated transmission matrix~\cite{Devaud2022TemporalLC}. For the spatial-only shaping results shown in Fig.~\ref{Fig2}, both spectral amplitude and phase are constant within the input bandwidth, so the incident pulse is transform-limited in time. We consider the time-gated TM $\mathfrak{T}_x(t)$ for the delay time $t=t^*$, and the eigenvector of $\mathfrak{T}_x(t)^\dagger \mathfrak{T}_x(t)$ with the largest eigenvalue gives the input spatial wavefront for the highest possible peak power that can be achieved with spatial-only shaping of a transform limited pulse. Dividing it by the peak transmitted power for random input wavefronts, $P_{\text{un}}(L,t^*)$, gives the enhancement by spatial wavefront shaping. Figure~\ref{Fig2}(c) shows that the peak enhancement decreases monotonically with increasing bandwidth $\Delta \nu$, as the lack of spectral control becomes more significant (SM Sec.~\ref{Sec:MP}~\cite{methods}). 
	
	\textbf {Spectral-only shaping}. For spectral-only wavefront shaping, the incident spatial wavefront $\psi_x(x_i)$ is preset, and the transmitted field is $\Psi_{\rm out}(x_o, t) = \int d \nu \, \mathfrak{T}_\nu(x_o, t; \nu) \, \psi_\nu(\nu)$, where $\mathfrak{T}_\nu(x_o, t; \nu) = \int d x_i \, e^{-i \, 2 \pi \, \nu t} \, T(x_o, x_i; \nu) \, \psi_x(x_i)$. In the matrix form, $\mathfrak{T}_\nu(t)$ is an $N_x \times N_\nu$ matrix, and $\psi_\nu(\nu)$ is a vector with $N_\nu$ elements. Similar to spatiospectral shaping, the maximum transmitted peak power is independent of the target time $t$, because the optimal pulse shape can be translated in time by a linear ramp in the input spectral phase. The peak enhancement with spectral-only shaping differs from spatial-only shaping by a factor $N_\nu/N_x$. In the current experiment, $N_x > N_\nu$; thus, controlling spatial channels provides higher enhancement (SM Sec.~\ref{Sec:MP}~\cite{methods}). 
	
	\textbf {Sequential spatial-spectral shaping}.	
	In spatial-only or spectral-only shaping, either the input spectrum or spatial wavefront is not optimized, leading to a reduced enhancement in the peak transmitted power. Therefore, we explore both spatial and spectral wavefront shaping by applying them sequentially to the incident pulse. The input spatiospectral wavefront can be written as $\Psi_{\rm in}(x_i, \nu) = \psi_x(x_i) \, \psi_\nu(\nu)$. We first fix the spatial wavefront $\psi_x(x_i)$ and optimize the spectrum $\psi_\nu(\nu)$. We then set the spectrum as the optimized one and optimize over the spatial wavefront. We iterate this process using experimentally measured transmission matrices until reaching convergence. 
	
	\bibliography{main}

    \newpage
    
    \section*{Supplementary Text}
    
    This document provides supplementary information to ``Generalized Time-Reversal for Pulse Control in Diffusive Media''. In the first section, we describe the experimental sample, the optical setup, and our measurement of the time-resolved spatiospectral transmission matrix (tSSTM). The second section details our numerical simulations. In the third section, we quantify the spectral channel width of our diffusive waveguide. The fourth section proves generalized time-reversal by deriving a relationship between the tSSTM and the broadband transmission matrix. The fifth section provides an analytic derivation of the spatiotemporal dynamics of an unmodulated transform-limited pulse inside a diffusive waveguide. In section six we derive the enhancement in the peak transmitted power using an effective Marchenko-Pastur model for spatiospectral, spectral-only, and spatial-only wavefront shaping. In the seventh section, we consider spatiotemporal focusing to a single spatial channel and analytically derive the spatiotemporal dynamics for pulse loading and firing. The eighth section covers the effect of dissipation on the loading and firing dynamics. Lastly, section nine explores time-reversal of loading and firing for deep pulse injection without dissipation.

    \section{Experimental Details}
	\label{Sec:Exp}
	
	Experiments are conducted in 2D disordered waveguides fabricated in a 220-nm thick silicon membrane on top of 3-\textmu m thick silica. At the probe wavelength $\sim$ 1550 nm, only the fundamental mode of TE polarization (field parallel to the membrane) is guided in the direction perpendicular to the silicon membrane. The 2D disordered waveguides are fabricated in the membrane and have sidewalls consisting of a triangular lattice of air holes with radius 155 nm and lattice constant 440 nm. The photonic-crystal boundary provides a full 2D photonic bandgap for TE light in the wavelength range of 1120-1580 nm. The disordered waveguide has width $W = 15$ \textmu m, length $L = 50$ \textmu m, and transport mean free path $\ell_t = 3.3$ \textmu m. It consists of randomly distributed air holes of 100 nm diameter and a filling fraction of 5.5\%. A weakly-scattering buffer zone, consisting of air holes with a filling fraction of 0.55\%, is placed in front of the disordered region to probe the incident field distribution. The transport mean free path $\ell_t = 33$ \textmu m exceeds the 25 \textmu m length of the buffer region, to reduce multiple scattering. Additional details on the sample parameters and fabrication can be found in Ref.~\cite{Bender2022}.

    \begin{figure*}
		\includegraphics[width=.65\textwidth]{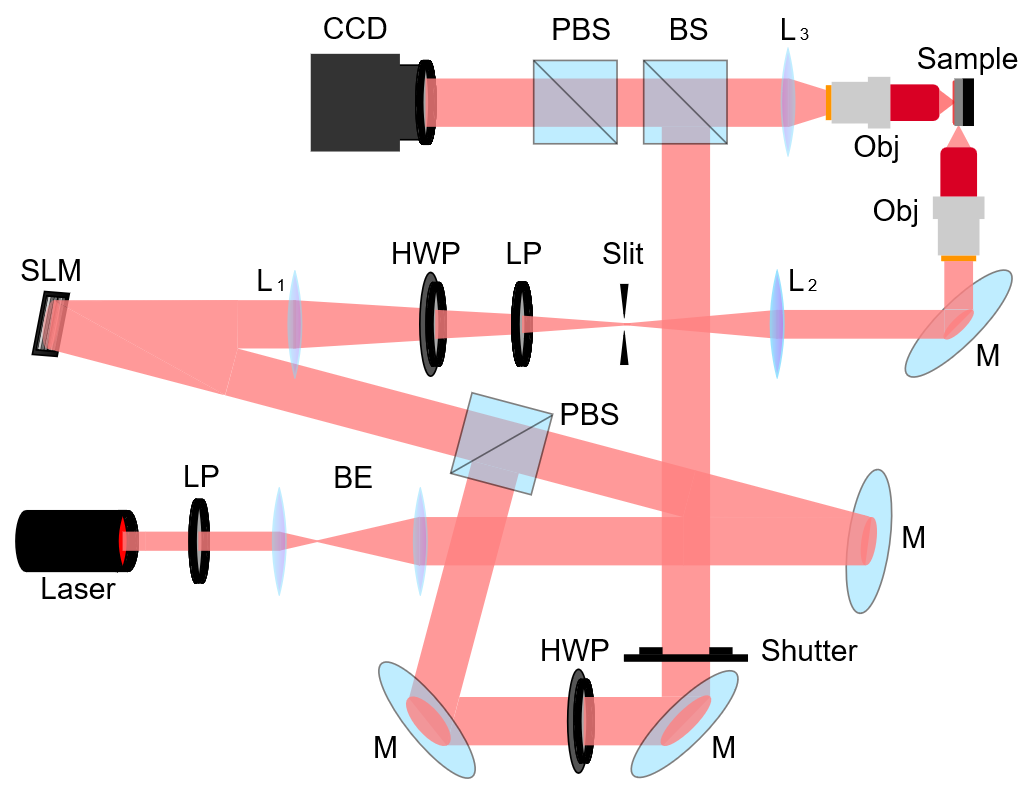}
		\caption{\textcolor{black}{\textbf{Experimental Setup.} A monochromatic, linearly-polarized beam from a tunable laser is split into two by a polarizing beam splitter (PBS). One beam is modulated by a spatial light modulator (SLM), then its polarization direction is rotated by a half-wave plate (HWP) before being launched to a planar waveguide in a silicon-on-insulator wafer through the side. The other beam serves as a reference beam and is combined with the out-of-plane scattered light from the waveguide using an unpolarized beam splitter (BS). After passing through a second PBS, the linearly-polarized interference pattern is recorded by a CCD camera. By varying the relative phase between the reference and the signal with the SLM, the field distribution inside the 2D disordered waveguide is recovered from the interference patterns. LP: linear polarizer; BE: beam expander; M: mirror; Obj: long-working-distance objective lens (Mitutoyo M Plan APO NIR HR100x, NA = 0.7); L: lens with focal length = 400 mm ($L_1$), 75 mm ($L_2$), and 100 mm ($L_3$). \label{setup}}}
	\end{figure*}
	
	\textcolor{black}{A detailed schematic of our optical setup is provided in Figure~\ref{setup}.} The monochromatic output of a tunable laser (Keysight 81960A) is split into a signal beam and a reference beam. The signal beam is modulated by a phase-only spatial light modulator (SLM, Hamamatsu LCoS X10468) and then injected into one of the optical waveguides via the edge of the wafer. The two lenses image the field reflected from the SLM onto the microscope objective pupil. As a result, the field injected into the waveguide corresponds to the two-dimensional Fourier transform of the field reflected from the SLM. A microscope objective (NA = 0.7) and a lens image the out-of-plane scattered optical field at the surface of the disordered waveguide onto a CCD camera chip (Allied Vision Goldeye G-032 Cool). The out-of-plane scattered light is then interfered with the reference beam on the CCD camera. To obtain the complex field distribution inside the sample, four phase shifts of $0$, $\pi/2$, $\pi$, and $3\pi/2$ are sequentially applied to the SLM, and the resulting interference patterns are recorded. The spatial resolution of our imaging system with NA $= 0.7$ is about $1.1$ \textmu m. 
		
	
	A linear mapping from the SLM to the field everywhere inside the sample, $T_{\text{SLM}\rightarrow \text{Int}}$, is measured as a function of frequency $\nu$. At each frequency, we apply orthogonal Hadamard patterns on $N_{\text{SLM}} = 128$ SLM macropixels and detect the field at all spatial locations inside the sample. The rank of this matrix is restricted to the number of waveguide modes $N_x<N_{\text{SLM}}$. In the probe wavelength range 1530 - 1575 nm, $N_x$ varies from 55 to 54.
	The transmission matrix is approximated by mapping to fields in a $10$-by-$10$ \textmu m$^2$ target region at the waveguide exit, $T_{\text{SLM}\rightarrow \text{Exit}}$. An extended region is chosen so that the number of speckles exceeds $N_x$. Due to the finite imaging resolution, the observed speckle size is around $1.1$ \textmu m, notably larger than the actual speckle size of  $W/N_x\approx 0.3$ \textmu m. The matrix $T_{\text{SLM}\rightarrow \text{Exit}}$ covers both light propagation from the SLM to the waveguide and that inside the disordered waveguide. To isolate light propagation in the disordered waveguide, we probe the field at the front entrance to the disordered waveguide by measuring the vertically scattered light from the buffer region. The transmission matrix from the field distribution in the buffer to that at the waveguide exit is then obtained using the Moore-Penrose pseudoinverse, $T_{\text{Buffer}\rightarrow \text{Exit}} = T_{\text{SLM}\rightarrow \text{Exit}}T_{\text{SLM}\rightarrow \text{Buffer}}^{+}$~\cite{Bender2022_2}. 
	
	The frequency-resolved transmission matrix $T_{\text{Buffer}\rightarrow \text{Exit}}(\nu) \equiv T(\nu)$ is measured by scanning the laser frequency $\nu$. Since the fields in the buffer and the disordered waveguide are measured with the same reference beam, the spectral phase between transmission matrices at different $\nu$ is obtained directly from the measurement. By horizontally concatenating frequency-resolved transmission matrices, we construct the time-resolved spatiospectral transmission matrix (tSSTM),
	\be
	\mathfrak{T}(t) = \frac{1}{\sqrt{\tilde{N}_\nu}} \left[ T(\nu_1)e^{-i2\pi\nu_1 t}, \dots, T(\nu_{\tilde{N}_\nu})e^{-i2\pi\nu_{\tilde{N}_\nu} t} \right],
	\label{EqDefSSTM1}
	\ee
	where $\tilde{N}_\nu$ is the number of frequency-resolved measurements (chosen to be larger than the number of uncorrelated spectral channels $N_\nu$, see Sec.~\ref{Sec:linewidth}). The tSSTM maps the spatial wavefronts at each frequency to the spatial profile in the target at a chosen time $t$. The input spatiospectral wavefront takes the form,
	\be
	\ket{\Psi_\text{in}} = \frac{1}{\sqrt{\tilde{N}_\nu}} \left[\ket{\psi_\text{in}(\nu_1)}, \dots, \ket{\psi_\text{in}(\nu_{\tilde{N}_\nu})} \right]^T,
	\label{EqSSWaveFront}
	\ee
	where $\ket{\psi_\text{in}(\nu_i)}$ is the input spatial wavefront at frequency $\nu_i$. The input is normalized to unity, $\sprod{\Psi_\text{in}}{\Psi_\text{in}}=1$. The eigenvector of $\mathfrak{T}(t)^\dagger\mathfrak{T}(t)$ with the largest eigenvalue maximizes the peak power at the waveguide exit at a target time $t$ (see Section~\ref{Sec:GTR} for details). With access to the mapping $T_{\text{SLM}\rightarrow \text{Int}}(\nu)$ at all frequencies, we recover the field distribution inside the waveguide as a function of frequency $\nu$ for a given input spatiospectral wavefront. The Fourier transform provides the internal field distribution as a function of time. This allows us to reconstruct the internal spatiotemporal dynamics for the dominant spatiospectral eigenchannel.
	
	Measurements of the tSSTM are performed in two wavelength ranges, 1530-1550 nm and 1555-1575 nm, with $N_x = 55$ and $N_x = 54$, respectively. The transition regime (1550-1555 nm) from $N_x = 55$ to $N_x = 54$ is avoided. The wavelength step is $d \lambda$ = 0.5 nm, therefore, the number of frequency-resolved measurements is $\tilde{N}_\nu=41$. A single measurement at 1555-1575 nm is used to reconstruct the loading and firing profile, while 9 repeated measurements at 1530-1550 nm provide the spatiospectral transmission eigenvalues. Repeated measurements improve the statistical estimation of the eigenvalues against experimental noise. The propagation of unmodulated pulses is reconstructed by initializing transform-limited pulses at different locations in the buffer region. The peak intensity at the waveguide exit is averaged over 100 pulses from different locations across the buffer.

    \textcolor{black}{The tSSTM allows us to reconstruct standard time-reversal experiments. To do this, we consider a random source field at the waveguide exit given by the column vector, $\vert\psi_\text{S}\rangle$. For a transform-limited pulse at the exit with this spatial wavefront, the transmitted spatiospectral wavefront in the buffer is given by $\mathfrak{T}(t)^T\vert\psi_\text{S}\rangle$. The time-reversed field in the buffer is acquired by phase-conjugating each frequency component. Therefore, the resulting spatiospectral input is $\mathfrak{T}(t)^\dagger\vert\psi_\text{S}\rangle^*$. All standard time-reversal (sTR) results are averaged over 100 random source wavefronts.}

    \textcolor{black}{For the deep pulse injection experiment, we use reciprocity to obtain the $R\to L$ (right to left) tSSTM from the measured $L\to R$ (left to right) matrices by transposition. The dominant singular vector of this reciprocal tSSTM gives the incident spatiospectral wavefront on the right that produces the fired pulse on the left. We then phase conjugate the transmitted field on the left and use the measured internal-field distribution to reconstruct the resulting $L\to R$ spatiotemporal power distribution inside the waveguide.}
	
	\section{Numerical Simulations}
	\label{Sec:Sim}
	
	Using KWANT~\cite{Groth2014}, we simulate disordered waveguides with reflecting sidewalls and compute the frequency-resolved transmission matrices. The waveguide parameters are length $L = 50, 100$ \textmu m, width $W = 15, 30$ \textmu m, and transport mean free path $\ell_t = 3.3, 6.6$ \textmu m. Dissipation is introduced by incorporating a diffusive absorption length $\xi_a = 28$ \textmu m~\cite{Yamilov2014}. The buffer region is neglected. 
	
	To simulate spatiotemporal dynamics, monochromatic transmission matrices are computed over a wavelength range of $\Delta\lambda = 20$ nm. For waveguides with $L = 50$ \textmu m and $L = 100$ \textmu m, we scan the wavelength $\lambda$ with step sizes of $d\lambda = 0.25$ nm and $d\lambda = 0.125$ nm respectively. Therefore, the number of sampled frequencies is $\tilde{N}_\nu = 81$. We then build the tSSTM using Eq.~\eqref{EqDefSSTM1}. To calculate the peak power enhancement, we compute the largest eigenvalue of $\mathfrak{T}(t)^\dagger\mathfrak{T}(t)$ and average over 100 realizations of disorder. The peak transmitted power is compared to that of transform-limited pulses with unmodulated wavefronts. The Thouless time $\tau_{\text{Th}} = L^2/\pi^2D$ gives the arrival time of light inside the sample, where $D$ is the diffusion constant. For a waveguide with $L = 50$ \textmu m and $\ell_t = 3.3$ \textmu m, $\tau_{\text{Th}} = 1.83$ ps is calculated by fitting the time-resolved transmitted power of an unmodulated pulse which decays as $e^{-t/\tau_\text{Th}}$ for times $t > \tau_\text{Th}$.
	
	Spatiotemporal intensity profiles are generated for single realizations of disorder. To resolve the field inside, a matrix mapping the input spatial wavefront to the field everywhere inside the waveguide is computed at each frequency~\cite{McIntosh2025}. Therefore, for a given spatiospectral wavefront as defined by Eq.~\eqref{EqSSWaveFront}, we can compute the spatial field profiles inside the waveguide at all frequencies. The Fourier transform then recovers the field inside as a function of time. Additional details on our numerical simulations can be found in Refs.~\cite{Sarma2014, Bender2020, McIntosh2025}.
	
	\section{Spectral Channel Width} \label{Sec:linewidth}
	
	The spectral channel width $\delta\nu$ determines the number of uncorrelated spectral channels $N_\nu$. Typically, $\delta\nu$ is estimated as the full-width-at-half-maximum (FWHM) of the short-range spectral correlation function~\cite{Hsu2015}, 
	\be
	C_1(|\nu_1 - \nu_2|) = \frac{\vert\overline{\Psi_{\rm out}(x_o, \nu_1) \Psi_{\rm out}^*(x_o, \nu_2)}\vert^2}{\overline{\vert\Psi_{\rm out}(x_o, \nu_1)\vert ^2}\ \overline{\vert\Psi_{\rm out}(x_o, \nu_2)\vert ^2}}.
	\label{EqDefC1}
	\ee
	$\overline{\vert{\Psi_{\rm out}(x_o, \nu)\vert}^2}$ is the transmitted intensity at location $x_o$ and frequency $\nu$, averaged over $x_o$, $\nu$, and random input wavefronts. This estimation, however, does not necessarily compute the exact number of uncorrelated channels $N_{\nu}$, which can be quantified directly from $C_1(|\nu - \nu_0|)$ by~\cite{McIntosh2024},
	\be
	\frac{1}{N_{\nu}}=\iint_{\Delta\nu}\frac{d\nu_1 d\nu_2}{\Delta\nu^2}C_1(\vert \nu_1-\nu_2 \vert),
	\label{EqMeffC1}
	\ee 
	for input pulses with a flat spectrum of width $\Delta\nu$.
	The spectral channel width is then obtained from $\delta\nu = \Delta\nu/(N_{\nu}-1)$ for $N_\nu >1$. 
	
	Equivalently, $N_{\nu}$ can be extracted from the spatial focusing of broadband light. First, we consider the transmission matrix at a single frequency, $T(x_o, x_i; \nu)$. Focusing the output at a single spatial location, $x_f$, is determined by a single row of $T(x_f, x_i; \nu)$. This gives the input spatial wavefront that optimizes focusing at $x_f$, and the enhancement of focal intensity over random input wavefronts is $N_x$. Next, to maximize the time-integrated power at the focus $x_f$ for a broadband input, we construct a broadband focusing matrix from $\tilde{N}_\nu$ frequency-resolved transmission matrices,
	\be
	\mathfrak{A}(x_f)=\frac{1}{\tilde{N}_\nu} \sum_{j=1}^{\tilde{N}_\nu} T(x_f, x_i; \nu_j)^\dagger T(x_f, x_i; \nu_j) \, .
	\label{EqBDM}
	\ee
	To ensure the transmission matrices are oversampled in frequency $\tilde{N}_\nu > N_\nu$, an initial estimation for $N_\nu$ can be made using the FWHM of $C_1(|\nu_1 - \nu_2|)$. The eigenvector of $\mathfrak{A}(x_f)$ with the largest eigenvalue is a single spatial wavefront that maximizes the time-integrated output power at the focus $x_f$ for a broadband input~\cite{Hsu2015, McIntosh2024}. For varying input bandwidth, the maximum focusing enhancement over random input wavefronts scales as $1/N_\nu$, where $N_\nu$ is the number of uncorrelated spectral channels, which does not depend on the degree of oversampling. Therefore, even with spectral oversampling in Eq.~\eqref{EqBDM}, Ref.~\cite{McIntosh2024} shows that $N_\nu$ can be directly computed from the variance of eigenvalues $\zeta$ of $\mathfrak{A}(x_f)$,
	\be
	\frac{1}{N_{\nu}}=\left\langle \frac{\text{Var}(\zeta)}{\text{Var}(\zeta_0)}\right\rangle \simeq\left\langle \frac{\text{Var}(\zeta)}{N_x\left\langle\zeta\right\rangle^2}\right\rangle,
	\label{EqVar}
	\ee
	where $N_x$ is the number of spatial channels, $\zeta_0$ are eigenvalues of the monochromatic focusing matrix with $N_\nu=1$, and the average is over $x_f$ and disorder realizations (more details in Sections III and V of the Supplementary Information in Ref.~\cite{McIntosh2024})
	
\begin{figure*}
		\includegraphics[width=0.8\textwidth]{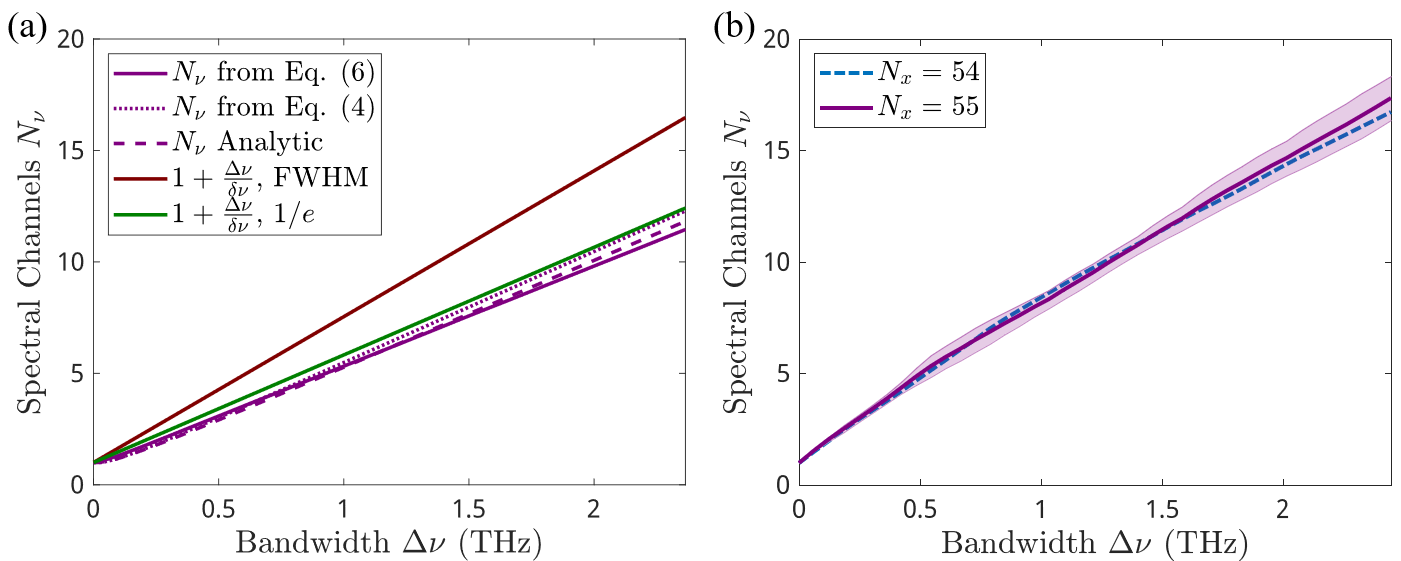}
		\caption{\label{FigSpecChan} \textbf{Number of spectral channels.} (a) Number of uncorrelated spectral channels $N_\nu$ as a function of the frequency bandwidth $\Delta \nu$, obtained from numerical simulations of frequency-resolved transmission matrices. $N_\nu$ calculated with Eq.~\eqref{EqVar} (purple solid) agrees with that from Eq.~\eqref{EqMeffC1}, using numerical (purple dotted) or analytical (purple dashed) $C_1(|\nu_1-\nu_2|)$. Taking the full width at $1/e$ of the maximum of $C_1(|\nu_1-\nu_2|)$ (green) as the spectral channel width gives the consistent $N_\nu$, while taking the FWHM of $C_1(|\nu_1-\nu_2|)$ (red) overestimates $N_\nu$. (b) $N_\nu$ obtained from experimentally measured frequency-resolved transmission matrices using Eq.~\eqref{EqVar} in two spectral ranges with $N_x = 54$ (cyan dashed) and $N_x = 55$ (purple). For $N_x = 55$, the purple shaded area marks the standard deviation over 9 repeated measurements.}
	\end{figure*}

	Figure~\ref{FigSpecChan}a shows that the $N_{\nu}$ values obtained from numerical simulations using Eq.~\eqref{EqMeffC1} and Eq.~\eqref{EqVar} are consistent. $C_1(|\nu_1 - \nu_2|)$ in Eq.~\eqref{EqMeffC1} can be computed either from numerical simulations or analytically (see Ref.~\cite{McIntosh2024} for a general expression and Eq.~\eqref{EqC1Transmission} for the result without absorption). We find that the FWHM of $C_1(|\nu_1 - \nu_2|)$ underestimates the spectral channel width, resulting in a larger $N_\nu$. Instead, taking the full width of $C_1(|\nu_1 - \nu_2|)$ at $1/e$ of the maximum as the spectral channel width leads to the correct channel number $N_\nu$. 
	
	For all experimental and numerical results presented here, Eq.~\eqref{EqVar} is used to calculate the number of uncorrelated spectral channels $N_{\nu}$. Figure~\ref{FigSpecChan}b shows the experimental $N_{\nu}$ values from frequency-resolved transmission matrices measured in two spectral ranges with $N_x = 54$ and $N_x = 55$, which are in good agreement. For $N_x = 55$, the purple band represents one standard deviation about the mean for 9 separate measurements. Note that the number of spectral channels measured in the experiment is larger than that from numerical simulation or analytic theory. The narrowing of the spectral channel width is attributed to measurement noise of the frequency-resolved transmission matrix. The experimentally measured matrices at varying frequencies contain random noise that accelerates spectral decorrelation, leading to an increase in $N_\nu$. For both the experiment and numerical simulations, the number of frequency-resolved measurements $\tilde{N}_\nu$ is chosen to be larger than the number of uncorrelated spectral channels, $\tilde{N}_\nu > N_\nu$
	
	\section{Generalized time-reversal}
	\label{Sec:GTR}
	
	We consider the general problem of optimizing the total transmission at a given time $t$ by controlling $N_\nu$ spectral channels and $N_1 \le N_x$ spatial channels of an incident pulse with bandwidth $\Delta \nu$. The case of $N_1<N_x$ represents incomplete spatial channel control. To formulate this as an eigenvalue problem, we write the transmitted field at time $t$ as
	\be
	\ket{\psi_\text{out}(t)} = \int_{\Delta \nu} \frac{d\nu}{\Delta \nu} e^{-i2\pi\nu t} T(\nu) \ket{\psi_\text{in}(\nu)},
	\label{EqFieldDepthL}
	\ee
	where $T(\nu)$ is the $N_x \times N_1$ transmission matrix at frequency $\nu$, and $\ket{\psi_\text{in}(\nu)}$ is the input state of spatial dimension $N_1$ and frequency $\nu$. The total transmission is then given by
	\begin{widetext}
		\begin{align}
			P_\text{out}(t) = \sprod{\psi_{\rm out}(t)}{\psi_{\rm out}(t)} &= \iint_{\Delta \nu} \frac{d\nu_1}{\Delta \nu} \frac{d\nu_2}{\Delta \nu} e^{-i2\pi(\nu_1 - \nu_2)t} \bra{\psi_\text{in}(\nu_2)} T(\nu_2)^\dagger T(\nu_1) \ket{\psi_\text{in}(\nu_1)}
			\nonumber \\
			&= \frac{1}{N_\nu^2} \sum_{i=1}^{N_\nu} \sum_{j=1}^{N_\nu} e^{-i2\pi(\nu_i - \nu_j)t} \bra{\psi_\text{in}(\nu_j)} T(\nu_j)^\dagger T(\nu_i) \ket{\psi_\text{in}(\nu_i)}.
			\label{EqOutputIntensity}
		\end{align}
	\end{widetext}
	We now reintroduce the $N_x \times N_\nu N_1$ tSSTM as
	\be
	\mathfrak{T}(t) = \frac{1}{\sqrt{N_\nu}} \left[ T(\nu_1)e^{-i2\pi\nu_1 t}, \dots, T(\nu_{N_\nu})e^{-i2\pi\nu_{N_\nu} t} \right],
	\label{EqDefSSTM}
	\ee
	whose frequency components satisfy
	\be
	\left[\mathfrak{T}(t)^\dagger \mathfrak{T}(t)\right]_{ji} = \frac{e^{-i2\pi(\nu_i - \nu_j)t}}{N_\nu} \, T(\nu_j)^\dagger T(\nu_i),
	\ee
	and redefine a generalized input state $\ket{\Psi_\text{in}}$ of size $N_\nu N_1$ as
	\be
	\ket{\Psi_\text{in}} = \frac{1}{\sqrt{N_\nu}} \left[\ket{\psi_\text{in}(\nu_1)}, \dots, \ket{\psi_\text{in}(\nu_{N_\nu})} \right]^T,
	\ee
	which is normalized to unity, $\sprod{\Psi_\text{in}}{\Psi_\text{in}}=1$.
	
	This allows us to rewrite Eq.~\eqref{EqOutputIntensity} as
	\be
	P_\text{out}(t) = \bra{\Psi_\text{in}} \mathfrak{T}(t)^\dagger \mathfrak{T}(t) \ket{\Psi_\text{in}},
	\label{EqIntensityTargetTime}
	\ee
	which immediately shows that the maximum transmission is achieved when the input state $\ket{\Psi_\text{in}}$ is the eigenvector $\ket{\Psi_1}$ of $\mathfrak{T}(t)^\dagger \mathfrak{T}(t)$ associated with its largest eigenvalue $\Lambda_1$.
	
	We also note that the simple time dependence of $\mathfrak{T}(t)$ implies that the eigenvalues of $\mathfrak{T}(t)^\dagger \mathfrak{T}(t)$ are independent of $t$, and that the eigenstates at any time $t$ can be directly deduced from those at $t=0$. The eigenvalue problem $\mathfrak{T}(t)^\dagger \mathfrak{T}(t)\ket{\Psi_n}=\Lambda_n \ket{\Psi_n}$, with $\ket{\Psi_n} = \left[\ket{\psi_n(\nu_1)}, \dots, \ket{\psi_n(\nu_{N_\nu})} \right]^T/\sqrt{N_\nu}$, reads
	\be
	\frac{1}{N_\nu}\sum_{i=1}^{N_\nu} e^{i2\pi\nu_j t} T(\nu_j)^\dagger T(\nu_i)e^{-i2\pi\nu_i t}\ket{\psi_n(\nu_i)}=\Lambda_n\ket{\psi_n(\nu_j)},
	\ee
	which becomes time-independent upon introducing $\ket{\tilde{\psi}_n(\nu_i)}= e^{-i2\pi\nu_i t}\ket{\psi_n(\nu_i)}$. This shows that the eigenvalues $\Lambda_n$ do not depend on $t$, and that the eigenstates at time $t$ are obtained from those at $t=0$ by multiplying each frequency component $i$ by a spectral phase factor $e^{i2\pi\nu_i t}$. Hence, without loss of generality, we can consider the optimization of total transmission at $t=0$. In the following, we use the simplified notation $\mathfrak{T}(t=0) \equiv \mathfrak{T}$.
	
	In order to understand the properties of the matrix $\mathfrak{T}^\dagger \mathfrak{T}$, it is instructive to make the connection with $\mathfrak{T}'^\dagger\mathfrak{T}'$, where $\mathfrak{T}' = \mathfrak{T}^T$. Explicitly, this matrix can be expressed as
	\be
	\mathfrak{T}'^\dagger\mathfrak{T}' = \frac{1}{N_\nu}\sum_{i=1}^{N_\nu} T'(\nu_i)^\dagger T'(\nu_i),
	\label{EqDefBroadbandTM}
	\ee
	where $T'(\nu_i) = T(\nu_i)^T$ is, according to reciprocity, the $N_1 \times N_x$ transmission matrix from right to left ($R\rightarrow L$) at frequency $\nu_i$. If we send a pulse with a single spatial wavefront $\ket{\psi'_\text{in}}$ of size $N_x$ from the right, the transmitted field to the left is
	\be
	\ket{\psi'(t)} = \int_{\Delta \nu} \frac{d\nu}{\Delta \nu} e^{-i2\pi\nu t} T'(\nu) \ket{\psi'_\text{in}}.
	\label{EqFieldDepthR2L}
	\ee
	The corresponding power $P'(t) = \sprod{\psi'(t)}{\psi'(t)}$ satisfies the property
	\be
	\int dt\, P'(t) = \frac{\bra{\psi'_\text{in}} \mathfrak{T}'^\dagger \mathfrak{T}' \ket{\psi'_\text{in}}}{\Delta \nu}.
	\label{EqTimeIntegrated}
	\ee
	This indicates that the time-integrated transmission in the $R\rightarrow L$ direction is maximized when sending the eigenstate $\ket{\psi'_1}$ associated with the largest eigenvalue $\Lambda'_1$ of the broadband transmission matrix $\mathfrak{T}'^\dagger \mathfrak{T}'$~\cite{Hsu2015}.
	
	Furthermore, we note that the non-zero eigenvalues of the matrix $\mathfrak{T}^\dagger \mathfrak{T}$, which are real and positive, are equal to those of the matrix $\mathfrak{T}'^\dagger\mathfrak{T}' = (\mathfrak{T}\mathfrak{T}^\dagger)^*$:
	\be
	\Lambda_n = \Lambda'_n.
	\label{EqEigenvalueEquality}
	\ee
	Combining this property with Eqs.~\eqref{EqIntensityTargetTime} and~\eqref{EqTimeIntegrated}, we find that the transmitted powers from $L\rightarrow R$ and from $R\rightarrow L$, resulting from the propagation of $\ket{\Psi_n}$ and $\ket{\psi'_n}$ respectively, satisfy the relation
	\be
	P_n(t = 0) = \Delta \nu \int dt\, P'_n(t).
	\label{EqRelationTransmittedIntensity}
	\ee
	This means that the peak transmitted power delivered by $\ket{\Psi_n}$ from $L\rightarrow R$, $P_n(t = 0)$, is equal to the time-integrated power $P'_n(t)$ delivered by the $n$-th eigenvector of the broadband transmission matrix $\ket{\psi'_n}$ from $R\rightarrow L$.
	

	A more general property relating propagation in the $R \to L$ and $L \to R$ directions can be identified by using the relation $\mathfrak{T}' = \mathfrak{T}^T$. This implies that the singular value decomposition of $\mathfrak{T}'$ can be written as
	\be
	\mathfrak{T}' = \sum_n \ket{\Psi_n^*} \Lambda_n^{1/2} \bra{\psi'_n},
	\ee
	where $\ket{\Psi_n}$ and $\ket{\psi'_n}$ are the eigenvectors of $\mathfrak{T}^\dagger\mathfrak{T}$ and $\mathfrak{T}'^\dagger\mathfrak{T}'$, respectively.
	This means that when the state $\ket{\psi'_n}$ is sent from the right, we collect the state $\ket{\Psi_n^*}$ on the left. Explicitly, each frequency component $\nu_i$ propagates from $R \to L$ according to
	\be
	T'(\nu_i)\ket{\psi'_n}=\sqrt{\Lambda_n} \ket{\psi_n(\nu_i)^*}.
	\ee
	This directly relates the eigenvectors of the tSSTM from $L \to R$ to those of the broadband transmission matrix from $R \to L$. Focusing on the dominant eigenvectors, $n = 1$, this implies that the transmitted power from $R \to L$ is
	\begin{align}
		P'_1(t) & =  \iint_{\Delta \nu} \frac{d\nu_1}{\Delta \nu} \frac{d\nu_2}{\Delta \nu} e^{-i2\pi(\nu_1 - \nu_2)t} \bra{\psi_1'}T'(\nu_2)^\dagger T'(\nu_1)\ket{\psi_1'}
		\nonumber \\
		& = \Lambda_1 \iint_{\Delta \nu} \frac{d\nu_1}{\Delta \nu} \frac{d\nu_2}{\Delta \nu} e^{-i2\pi(\nu_1 - \nu_2)t} \sprod{\psi_1(\nu_2)}{\psi_1(\nu_1)}^*
		\nonumber \\
		&= \Lambda_1 \iint_{\Delta \nu} \frac{d\nu_1}{\Delta \nu} \frac{d\nu_2}{\Delta \nu} e^{i2\pi(\nu_1 - \nu_2)t} \sprod{\psi_1(\nu_2)}{\psi_1(\nu_1)}
		\nonumber \\
		&= \Lambda_1 P_{1}^\text{in}(-t),
		\label{EqLRTimeReversal}
	\end{align}
	where $P_{1}^\text{in}(t)$ is the power associated with the field $\ket{\Psi_1}$. Hence, the transmitted power $P'_1(t)$ at time $t$ from $R \to L$ resulting from the propagation of $\ket{\psi'_1}$ is proportional to the time-reversal of the input power $P_{1}^\text{in}(-t)$ of the state $\ket{\Psi_1}$ that maximizes peak transmission from $L \to R$. Integrating Eq.~\eqref{EqLRTimeReversal} over time, with $\Lambda_1=P_1(t=0)$ [see Eq.~\eqref{EqIntensityTargetTime}] and the normalization of the input states ($\Delta \nu \int dt\, P_{1}^\text{in}(t) = 1$), we recover Eq.~\eqref{EqRelationTransmittedIntensity}.


	\section{Unmodulated Pulse Propagation} \label{Sec:Unmodulated}
	
	Since the transmission enhancement is measured relative to that of an unmodulated pulse, we analytically study the propagation of a transform-limited pulse with a uniform input wavefront through the diffusive medium. The input pulse has a flat spectrum of bandwidth $\Delta \nu$ and is launched at $t=0$. Its spatial wavefront is given by $|\psi_\text{in} \rangle = 1/\sqrt{N_x}$, where $N_x$ is the number of spatial channels. The intensity integrated along the transverse direction of the waveguide is $P_\text{in}(t) = \sprod{\psi_\text{in}(t)}{\psi_\text{in}(t)} = \,\text{sinc}^2(\pi \Delta \nu t)$, where the Fourier transform $\ket{\psi_\text{in}(t)}=\int_{\Delta \nu} e^{-i2\pi\nu t} \ket{\psi_\text{in}(\nu)} {d\nu}/{\Delta \nu}$ represents the field in the time domain. 
	
	To find the internal field distribution at depth $0\le z\le L$ and time $t$, we calculate,
	\be
	\ket{\psi(z,t)} = \int_{\Delta \nu} \frac{d\nu}{\Delta \nu} e^{-i2\pi\nu t} \mathcal{Z}(z,\nu) \ket{\psi_\text{in}(\nu)},
	\label{EqFieldDepthZ}
	\ee
	where $\mathcal{Z}(z,\nu)$ is the $N_x \times N_x$ deposition matrix that maps the incident wavefront to the field across the waveguide cross-section at depth $z$~\cite{Bender2022_2}. The deposition matrix enables a prediction of the field everywhere inside the medium, which is used later in Sec.~\ref{Sec:GTRvsTR}. For $z = L$, the deposition matrix reduces to the transmission matrix, $\mathcal{Z}(L,\nu) = T(\nu)$. The resulting mean power at depth $z$ and time $t$ becomes
	\begin{align}
		&P_\text{un}(z,t) = \overline{\sprod{\psi(z,t)}{\psi(z,t)}}
		\nonumber\\
		&= \iint_{\Delta \nu} \frac{d\nu_1}{\Delta \nu} \frac{d\nu_2}{\Delta \nu} \frac{1}{N_x}\sum_{i,j,k}^{N_x} e^{-i2\pi(\nu_1-\nu_2)t}
		\overline{ \mathcal{Z}_{ji}(z,\nu_1)\mathcal{Z}_{jk}(z,\nu_2)^* }
		\nonumber\\
		&= \bar{\zeta}(z) \mathcal{F}(z,t),
		\label{EqMeanIntensity}
	\end{align}
	where $\overline{\cdots}$ represents averaging over realizations of disorder. Here, we make use of the property $\overline{ \mathcal{Z}_{ji}(z,\nu_1)\mathcal{Z}_{jk}(z,\nu_2)^* } = \overline{ \mathcal{Z}_{ji}(z,\nu_1)\mathcal{Z}_{ji}(z,\nu_2)^*} \delta_{ik}$, and introduce the mean power at depth $z$, $\bar{\zeta}(z) = \overline{\text{Tr}\left[\mathcal{Z}(z,\nu)^\dagger \mathcal{Z}(z,\nu)\right]}/N_x$. 
	The latter reads
	\begin{align}
		\bar{\zeta}(z) &= 2\left[1-\bar{T}\text{cosh}\left(\frac{L}{\xi_a} \right)\right]\frac{\text{sinh}\left(\frac{L-z}{\xi_a} \right)}{\text{sinh}\left(\frac{L}{\xi_a} \right)}
		\nonumber
		\\
		&+ \bar{T}\text{cosh}\left(\frac{L-z}{\xi_a}\right),
		\label{EqMeanIntensityZAbs}
	\end{align}
	where $\bar{T} = \bar{\zeta}(L)\simeq (\pi \ell_t/2\xi_a) [\text{sinh}(L/\xi_a)+(\pi \ell_t/2\xi_a)\text{cosh}(L/\xi_a)]^{-1}$. To obtain the result~\eqref{EqMeanIntensityZAbs}, we generalized the approach presented in Section 2.2 of the Supplementary Information of Ref.~\cite{Bender2022_2} to include absorption, using expression (27) from that reference for the mean power and assuming moderate absorption ($\xi_a \gg \ell_t$).
	The quantity $\mathcal{F}(z,t)$ in Eq.~\eqref{EqMeanIntensity} is the arrival-time distribution defined as
	\begin{align}
		\mathcal{F}(z,t) &= \iint_{\Delta \omega} \frac{d\omega_1}{\Delta \omega} \frac{d\omega_2}{\Delta \omega} e^{-i(\omega_1-\omega_2)t} C_E(z, \omega_1-\omega_2)
		\nonumber\\
		&= 2\,\text{Re}\left[\int_{\Delta \omega} \frac{d\Omega}{\Delta\omega} \frac{\Delta\omega - \Omega}{\Delta\omega} e^{-i\Omega t} C_E(z,\Omega)\right].
		\label{EqDwellTime}
	\end{align}
	Here and in the following, we use the angular frequency $\omega = 2\pi \nu$ instead of $\nu$. The spectral bandwidth is $\Delta\omega = 2\pi \Delta\nu$, and $C_E$ is the field correlation function, defined as $C_E(z,\Omega)=\overline{\mathcal{I}(z,\Omega)}/\overline{\mathcal{I}(z,0)}$, with $\overline{\mathcal{I}(z,\Omega)} =\overline{\mathcal{Z}_{ji}(z,\omega +\Omega/2)\mathcal{Z}_{ji}^*(z,\omega -\Omega/2) }$ and $\mathcal{Z}_{ji}(z,\omega)$ giving the field at depth $z$. The correlator, independent of the indices $i$ and $j$, can be expressed as $\overline{\mathcal{I}(z,\Omega)}=\int_0^Ldz'e^{-z'/\ell_t}K(z,z',\Omega)$, where $K(z,z',\Omega)$ is the Green's function of the diffusion equation
	\be
	\left[\left(-\partial^2_z+\frac{1}{\xi^2_a}\right)-\frac{i\Omega}{D}\right]K(z,z',\Omega) = \delta(z-z'),
	\ee
	with boundary conditions $\partial_z K(0,z',\Omega) = K(0,z',\Omega)/z_0$ and $\partial_z K(L,z',\Omega) = -K(L,z',\Omega)/z_0$, where $z_0=\pi \ell_t/4$ is the extrapolation length and $D$ is the diffusion constant. Introducing $L_\Omega=\sqrt{D/2\Omega}$, the coherence length of the diffusive light with frequency detuning $\Omega$, we obtain, for $\ell_t \ll L, L_\Omega$, 
	\begin{widetext}
		\be
		C_E(z, \Omega) = \frac{\text{sinh}\left(\frac{L}{\xi_a} \right)}{\text{sinh}\left(\frac{L-z}{\xi_a} \right)}\ \frac{\text{cos}\left(\beta\frac{\tilde{L}-\tilde{z}}{2}\right)\text{sinh}\left(\alpha\frac{\tilde{L}-\tilde{z}}{2}\right)-i\,\text{sin}\left(\beta\frac{\tilde{L}-\tilde{z}}{2}\right)\text{cosh}\left(\alpha\frac{\tilde{L}-\tilde{z}}{2}\right)}{\text{cos}\left(\beta\frac{\tilde{L}}{2}\right)\text{sinh}\left(\alpha\frac{\tilde{L}-\tilde{z}}{2}\right)-i\,\text{sin}\left(\beta\frac{\tilde{L}-\tilde{z}}{2}\right)\text{cosh}\left(\alpha\frac{\tilde{L}-\tilde{z}}{2}\right)},
		\label{EqCE}
		\ee
	\end{widetext}
	where $\tilde{L}=L/L_\Omega$ and $\tilde{z}=z/L_\Omega$, and 
	\begin{align}
		\alpha &=\left[\left(1+\frac{4}{\tilde{\xi_a}^4}\right)^{1/2}+\frac{2}{\tilde{\xi_a}^2}\right]^{1/2}, 
	\end{align}
	\begin{align}
		\beta &=\left[\left(1+\frac{4}{\tilde{\xi_a}^4}\right)^{1/2}-\frac{2}{\tilde{\xi_a}^2}\right]^{1/2},
	\end{align}
	with $\tilde{\xi_a}=\xi_a/L_\Omega$. The arrival-time distribution $\mathcal{F}(z,t)$ follows from the combination of Eqs.~\eqref{EqDwellTime} and~\eqref{EqCE}.
	
	\begin{figure}[t]
		\includegraphics[width=0.45\textwidth]{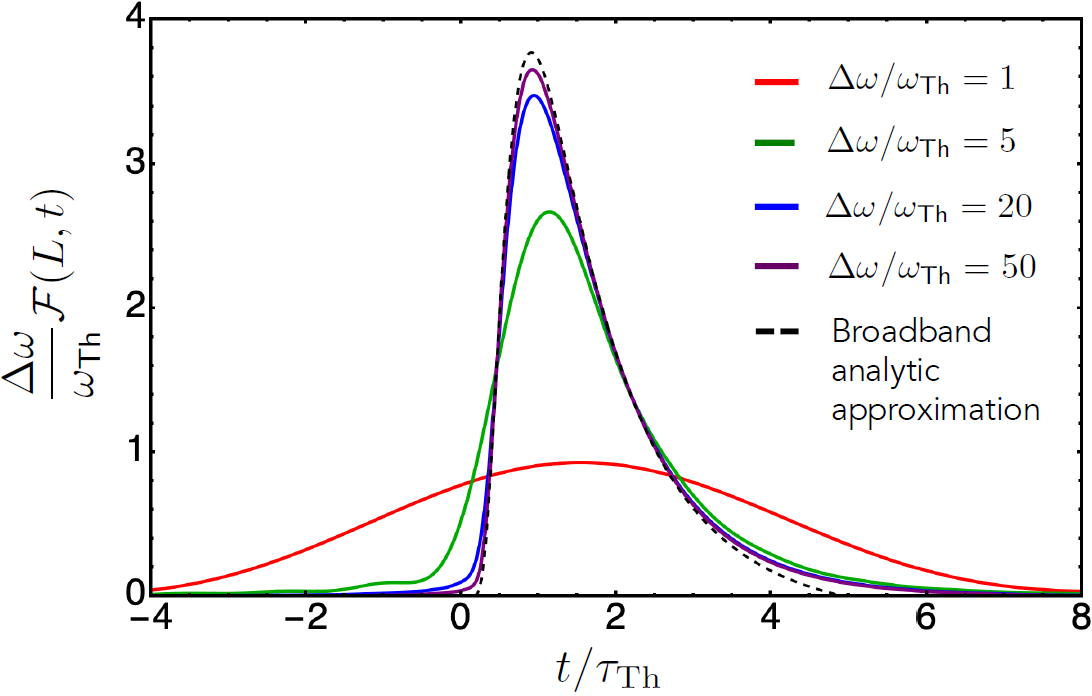}
		\caption{\label{FigArrivalTime} 
			\textbf{Arrival-time distribution.} A transform-limited pulse with a random input wavefront is launched to a diffusive medium at $t=0$. The pulse has a flat spectrum of bandwidth $\Delta \omega =2\pi \Delta \nu$ normalized by the Thouless frequency $\omega_{\text{Th}}=\pi^2 D/L^2$. Theoretically calculated distribution $\mathcal{F}(L,t)=P_\text{un}(L,t)/\bar{T}$ [see Eq.~\eqref{EqDwellTime}] as a function of the arrival time $t$. The broadband analytic approximation (dashed line) is given by Eq.~\eqref{EqApproxArrival}.
		}
	\end{figure}
	
	In the broadband limit $\Delta \omega \gg \omega_{\text{Th}} = \pi^2 D/L^2$ and without absorption, we find that the arrival-time distribution is well approximated by
	\be
	\mathcal{F}(z,t) \simeq \frac{\pi^{3/2}}{ \Delta \tilde{\omega} \tilde{t}^{3/2}} \frac{L}{L-z} \left[
	\frac{z}{L} e^{-\frac{\pi^2}{4 \tilde{t}} \frac{z^2}{L^2}} - \frac{2L-z}{L} e^{-\frac{\pi^2}{4 \tilde{t}} \frac{(2L-z)^2}{L^2}}
	\right],
	\ee
	where $\Delta \tilde{\omega} = \Delta \omega/\omega_{\text{Th}}$, and the time is expressed in units of the Thouless time $\tau_{\text{Th}} = 1/\omega_{\text{Th}}$ as $\tilde{t} = t/\tau_{\text{Th}}$. In particular, for transmission at $z=L$, this simplifies to
	\be
	\mathcal{F}(L,t) \simeq \frac{\pi^{3/2}(\pi^2-2\tilde{t}) e^{-\frac{\pi^2}{4 \tilde{t}}}}{ \Delta \tilde{\omega} \,\tilde{t}^{5/2}}.
	\label{EqApproxArrival}
	\ee
	This broadband approximation is compared to the exact result~\eqref{EqDwellTime} in Fig.~\ref{FigArrivalTime}.
	We note that it reaches the maximum at a time close to the Thouless time,
	\be
	t^* = \left(\frac{3-\sqrt{6}}{6}\pi^2\right) \tau_{\text{Th}} \simeq 0.906 \, \tau_{\text{Th}},
	\label{EqArrivalTime}
	\ee
	which gives
	\be
	\mathcal{F}(L,t^*) = \frac{72e^{-(3+\sqrt{3})/2}}{(3-\sqrt{6})^{5/2}\pi^{3/2}}\frac{1}{\Delta \tilde{\omega}} \simeq \frac{3.77}{\Delta \tilde{\omega}}.
	\ee
	Since in the broadband limit we also have $N_\nu = 1/\bar{C}_1(L,\Delta \omega) \simeq \pi^2 \Delta \tilde{\omega}/24$ [see Eqs.~\eqref{EqMeffC1},~\eqref{EqDefCk} and~\eqref{EqC1Transmission}], we conclude that the peak transmission of a transform-limited pulse with a random input wavefront scales as
	\be
	\text{max}_t[P_{\text{un}}(L,t)] = \bar{T} \,\mathcal{F}(L,t^*) \simeq \frac{1.5\, \bar{T}}{N_\nu}.
	\ee
	We note that the above predictions for $t^*$ and $\mathcal{F}(L,t^*)$ remain nearly unchanged in the presence of moderate absorption ($\xi_a\gg\ell_t$), such as that considered in our experiment.
	
    \section{Effective Marchenko-Pastur model}
    \label{Sec:MP}
	
	\subsection{Spatiospectral Shaping}
	
	We establish in Eq.~\eqref{EqEigenvalueEquality} that the non-zero eigenvalues of the tSSTM are identical to those of the reciprocal broadband matrix $\mathfrak{T}'^\dagger \mathfrak{T}'$ introduced in Eq.~\eqref{EqDefBroadbandTM}, where $T'(\nu)$ are transmission matrices in the $R\to L$ direction of size $N_1 \times N_x$. The case of $N_1 = N_x$ represents full input spatial channel control while $N_1<N_x$ represents incomplete spatial channel control. The eigenvalue distribution of the broadband transmission matrix has been studied in detail in Refs.~\cite{Hsu2015, McIntosh2024}. In particular, it was demonstrated that the full eigenvalue spectrum is well captured by the filtered random matrix (FRM) theory originally introduced in Ref.~\cite{Goetschy2013}, with renormalized filtering parameters to account for broadband long-range correlations. While the FRM theory is accurate, it does not provide explicit closed-form expressions for the eigenvalue spectrum.
	
	A less accurate but more intuitive approach is the effective Marchenko-Pastur model~\cite{McIntosh2024}, which
    amounts to assuming that the matrix $\mathfrak{T}'^\dagger \mathfrak{T}'$, where $\mathfrak{T}'$ is of size $ N_\nu  N_1\times N_x$, can be approximated by a Wishart matrix $H^\dagger H$, with $H$ a Gaussian random matrix of size $N_\text{eff} \times N_x$. The effective number of degrees of freedom, $N_\text{eff}$, is smaller than $N_1 N_\nu$ due to long-range spatial and spectral correlations on the left side. As a result, this model yields explicit expressions for the eigenvalue density in terms of the short- and long-range broadband contributions of the speckle patterns, denoted $\bar{C}_1$ and $\bar{C}_2$, respectively. They are defined as
	\be
	\bar{C}_{k}=\iint_{\Delta\omega}\frac{d\omega_1d\omega_2}{\Delta\omega^2}C_{k}(|\omega_1-\omega_2|),
	\label{EqDefCk}
	\ee
	where $C_{1}(\Omega)$ and $C_{2}(\Omega)$ are the two contributions to the total transmission correlation function, $\mathcal{C}(\Omega)=\overline{\mathcal{T}(\omega + \Omega/2) \mathcal{T}(\omega-\Omega/2) }/\overline{\mathcal{T}(\omega) }^2-1$, where  $\mathcal{T}(\omega)=\sum_j \vert T_{ji}(\omega)\vert^2$~\cite{Akkermans2007}. In the absence of absorption, they are
	\begin{align}
		C_1(\Omega)&=\frac{\tilde{L}^2}{\text{cosh}(\tilde{L})-\text{cos}(\tilde{L})},
		\label{EqC1Transmission}
		\\
		C_2(\Omega)&=\frac{2}{g} \frac{1}{\tilde{L}}\frac{\text{sinh}(\tilde{L})-\text{sin}(\tilde{L})}{\text{cosh}(\tilde{L})-\text{cos}(\tilde{L})},
		\label{EqC2Transmission}
	\end{align}
	where $g=N_x\bar{T}$ and $\tilde{L}=L/L_\Omega$, with $L_\Omega=\sqrt{D/2\Omega}$. Ref.~\cite{McIntosh2024} provides a generalization of these expressions to arbitrary depth and in the presence of absorption. 
	
	The effective Marchenko-Pastur model expresses the number of independent channels $N_\text{eff}$ contributing to the time-integrated total transmission in terms of the number of output channels, as well as $\bar{C}_1$ and $\bar{C}_2$. The corresponding effective Wishart matrix's eigenvalues $\Lambda' $are parametrized by $N_\text{eff}$, $N_x$, and their mean $\overline{\Lambda'}$. We refer the reader to Section VI of the Supplementary Material (SM) of Ref.~\cite{McIntosh2024} for more details.
	
	Since the effective Marchenko-Pastur model yields reasonable agreement with the FRM theory for the largest eigenvalue---which is the main focus of the present section---we will discuss the properties of the largest tSSTM eigenvalue using this simplified formalism. As noted in Sec.~\ref{Sec:GTR}, the mapping with the tSSTM applies to the reciprocal BTM describing broadband transmission from $R \to L$. Therefore, by taking $N_x$ as the number of input channels and $N_1$ as the number of output channels, and by selecting the depth $z=L$ in the predictions established in Section VI of the SM of Ref.~\cite{McIntosh2024}, we find that the eigenvalues $\Lambda'$ of $\mathfrak{T}'^\dagger \mathfrak{T}'$ satisfy
	\be
	\overline{\text{max} \, \Lambda'} = \overline{\Lambda'} \left(1 + \sqrt{\frac{N_x}{N_\text{eff}}} \right)^2,
	\ee
	where $\overline{\Lambda'} = (N_1 / N_x) \bar{T}$ and  
	$1/N_\text{eff} = \bar{C}_1 / N_1 + \bar{C}_2$. Note that $N_\text{eff}$ is also related to the variance of the eigenvalues $\Lambda'$~\cite{McIntosh2024}:
	\be
	\frac{1}{N_\text{eff}} = \frac{1}{N_x} \left( \frac{\overline{\Lambda'^2}}{\overline{\Lambda'}^2} - 1 \right).
	\ee
	In addition, according to Eqs.~\eqref{EqIntensityTargetTime} and~\eqref{EqEigenvalueEquality}, the peak transmission is given by $\text{max}_t[P(L,t)] = \overline{\text{max} \, \Lambda} = \overline{\text{max} \, \Lambda'}$. Normalizing this by the peak transmission of an unmodulated pulse studied in Sec.~\ref{Sec:Unmodulated}, $\text{max}_t[P_\text{un}(L,t)] =  \bar{T} \, \mathcal{F}(L,t^*)$, we obtain the peak transmission enhancement:
	\begin{widetext}
		\be
		\frac{\text{max}_t[P(L,t)]}{\text{max}_t[P_\text{un}(L,t)]} = \frac{N_1}{N_x} \frac{1}{\mathcal{F}(L,t^*)} \frac{\overline{\text{max} \, \Lambda'}}{ \overline{\Lambda'} }
		= \left(
		\sqrt{\frac{N_1}{N_x} \frac{1}{\mathcal{F}(L,t^*)}} + \sqrt{ \frac{\bar{C}_1}{\mathcal{F}(L,t^*)} + \frac{N_1 \bar{C}_2}{\mathcal{F}(L,t^*)} }
		\right)^2.
		\label{EqEnhancementArbitrarySize}
		\ee
	\end{widetext}
	
\begin{figure*}
		\includegraphics[width=0.8\textwidth]{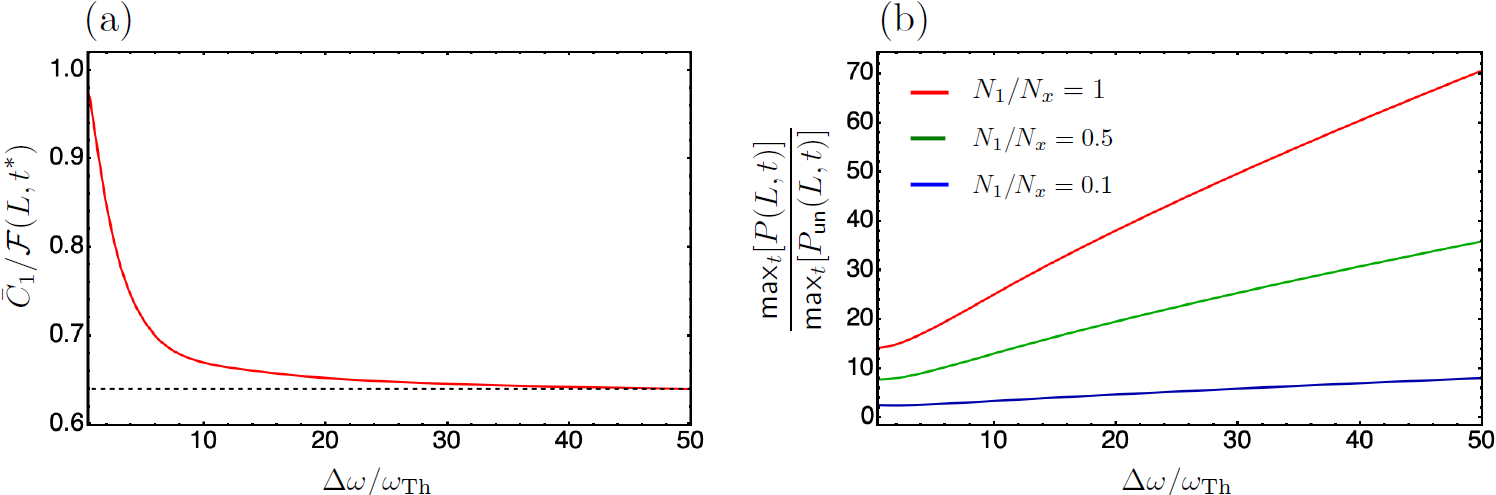}
		\caption{\label{FigMPModel} \textbf{Theoretical prediction for peak transmission enhancement.} (a) Ratio between the short-range correlation contribution $\bar{C}_1 = 1/N_\nu$ and the peak of the arrival-time distribution $\mathcal{F}(L,t^*)$ in Eq.~\eqref{EqEnhancementArbitrarySize}. This ratio remains below unity and thus contributes only weakly to the transmission enhancement. In the broadband limit, it tends to $\frac{(3 - \sqrt{6})^{5/2}}{3\sqrt{\pi}} e^{\frac{3}{2} + \sqrt{\frac{3}{2}}} \simeq 0.64$ (dashed line). (b) Transmission enhancement at the optimal time $t^*$ through a diffusive medium with mean transmission $\bar{T} = 0.1$, as predicted by Eq.~\eqref{EqEnhancementArbitrarySize}, for different numbers $N_1$ of input channels. Note that in the broadband limit, the normalized bandwidth $\Delta \omega / \omega_{\text{Th}}$ is simply proportional to $N_\nu$, with the relation $N_\nu \simeq (\pi^2 / 24) \, \Delta \omega / \omega_{\text{Th}} \simeq 0.41 \, \Delta \omega / \omega_{\text{Th}}$.}
		\label{Fig:MPmodel} 
	\end{figure*}

	The expression~\eqref{EqEnhancementArbitrarySize} contains three distinct contributions. The first term dominates when focusing to a small number of output channels ($N_x$ is replaced by $N_2 \ll N_1$). Since $\mathcal{F}(L,t^*) \propto 1/N_\nu$, the corresponding enhancement scales as $N_1 N_\nu / N_2$. As shown in Fig.~\ref{Fig:MPmodel}(a), the second contribution, $\bar{C}_1 / \mathcal{F}(L,t^*)$, is always less than unity and converges to 0.64 for $\Delta \omega \gg \omega_{\text{Th}}$. As discussed in Sec.~\ref{Sec:Unmodulated}, we have $\mathcal{F}(L,t^*) \simeq 3.77\  \omega_{\text{Th}} / \Delta \omega$ and $\bar{C}_1 \simeq (24 / \pi^2) \omega_{\text{Th}} / \Delta \omega$ in this limit. 
	The third term in Eq.~\eqref{EqEnhancementArbitrarySize}, associated with the long-range spectral correlations, becomes the dominant contribution when $N_1$ is large and approaches $N_x$. In the broadband limit, where $\bar{C}_2 \simeq (32 / 3\sqrt{2} \pi^2)(L / N_x \ell_t) \sqrt{ \omega_{\text{Th}} / \Delta \omega }$~\cite{McIntosh2024}, this term scales as
	\be
	\frac{N_1 \bar{C}_2}{\mathcal{F}(L,t^*)} \simeq \frac{0.5}{\bar{T}} \frac{N_1 \sqrt{N_\nu}}{N_x}.
	\label{EqLeadingEnhancement}
	\ee
	In the case of full spatial control, $N_1 = N_x$, Eq.~\eqref{EqEnhancementArbitrarySize} becomes,
	\be
	\frac{\text{max}_t[P(L,t)]}{\text{max}_t[P_\text{un}(L,t)]}
	= \frac{1}{\mathcal{F}(L,t^*)}\left(
	1 + \sqrt{ \bar{C}_1+N_x\bar{C}_2}
	\right)^2.
	\label{EqEnhancementCompleteControl}
	\ee
	
	In Fig.~\ref{Fig:MPmodel}(b), we show the prediction of Eq.~\eqref{EqEnhancementArbitrarySize}, evaluated using Eqs.~\eqref{EqDwellTime},~\eqref{EqCE},~\eqref{EqDefCk},~\eqref{EqC1Transmission} and~\eqref{EqC2Transmission}, as a function of the bandwidth for various values of the number of spatial channels $N_1$.
	In the main text, this prediction, evaluated with the generalization of Eqs.~\eqref{EqC1Transmission} and~\eqref{EqC2Transmission} to the diffusive system with absorption, as given in Ref.~\cite{McIntosh2024},  is shown to be in good agreement with both numerical simulations and experimental results, for the case of full spatiospectral control ($N_1 = N_x$).
	
	\subsection{Spectral-Only Shaping}
	
	The effective Marchenko-Pastur model also predicts the peak power enhancement for spectral-only wavefront shaping. In this case, the input pulse has a fixed spatial wavefront, assumed here to be random, and the spectral pulse shape can be modulated. This is the special case of spatiospectral wavefront shaping where $N_1 = 1$. As a result, $1/N_\text{eff} = \bar{C}_1 + \bar{C}_2$. Equation~\eqref{EqEnhancementArbitrarySize} then reduces to
	\be
	\frac{\text{max}_t[P(L,t)]}{\text{max}_t[P_\text{un}(L,t)]}
	= \frac{1}{N_x\mathcal{F}(L,t^*)}\left(
	1 + \sqrt{N_x\bar{C}_1+N_x\bar{C}_2}
	\right)^2.
	\label{EqEnhancementSpecOnly}
	\ee
	For large input bandwidths, this gives a $1/N_x$ reduction in the enhancement compared to the case of spatiospectral wavefront shaping with full spatial control Eq.~\eqref{EqEnhancementCompleteControl}. Eq.~\eqref{EqEnhancementSpecOnly} agrees with our numerical and experimental results shown in Fig. 2 of the main text.
	
	\subsection{Spatial-only Shaping}
	
	When only spatial control is accessible, we consider a transform-limited input pulse where the input spatial wavefront is the same for all frequencies, $\ket{\psi_{\text{in}}(\nu_i)}=\ket{\psi_{\text{in}}}$. The total transmitted power at time $t$, Eq.~~\eqref{EqOutputIntensity}, can then be written as
	\be
	P(t)=\bra{\psi_{\text{in}}}T(t)^\dagger T(t) \ket{\psi_{\text{in}}},
	\ee
	where $T(t)$ is the time-gated transmission matrix,
	\be
	T(t) = \int_{\Delta \nu} \frac{d\nu}{\Delta \nu} e^{-i2\pi\nu t} T(\nu).
	\ee
	In this situation, the largest transmission at time $t$ is achieved by considering the eigenstate of $T(t)^\dagger T(t)$ associated with its largest eigenvalue $\sigma_1(t)$.
	
	In the framework of the effective Marchenko-Pastur approach, the $N_x \times N_1$ matrix $T(t)$ is replaced by a Gaussian random matrix of dimension $N_\text{eff}(t) \times N_1$, where $N_\text{eff}(t)$ is evaluated by computing the number of independent speckles in the output target at time $t$, for a single-channel excitation. Denoting $\mathcal{T}(t)=\sum_{j=1}^{N_x} \vert T_{ji}(t) \vert^2$ as the output transmission when only spatial channel $i$ is excited, we define $N_\text{eff}(t)$, independent of $i$, as
	\be
	\frac{1}{N_\text{eff}(t)} = \frac{\overline{\mathcal{T}(t)^2}}{\overline{\mathcal{T}(t)}^2}-1.
	\label{EqDefNeffTime}
	\ee
	This implies that $\mathcal{T}(t)$ can be represented as $\mathcal{T}(t)=\sum_{j=1}^{N_\text{eff}(t)}\tilde{\mathcal{T}}_{ji}(t)$, where $\tilde{\mathcal{T}}_{ji}(t)$ is the transmission coefficient. By singular value decomposition of the matrix $T(t)$, it can be shown that, for $N_1\gg 1$, this definition is equivalent to 
	\be
	\frac{1}{N_\text{eff}(t)} = \frac{1}{N_1} \left(\frac{\overline{[\sigma(t)]^2}}{[\overline{\sigma(t)}]^2}-1\right),
	\label{EqDefNeffTime2}
	\ee
	where $\overline{[\sigma(t)]^k}$ denotes the $k$-th moment of the distribution $p(\sigma, t)$ of the eigenvalues $\sigma_n(t)$ of the matrix  $T(t)^\dagger T(t)$: $\overline{\sigma(t)^k}=\int d\sigma p(\sigma, t)\sigma^k $.
	A detailed proof of this relation can be found by combining the results of the supplementary section 2.2 of Ref.~\cite{Bender2022} and the supplementary section 2.B of Ref.~\cite{Bender2022_2}.
	
	Equation~\eqref{EqDefNeffTime} can be evaluated explicitly using the following expansion of the speckle correlation function:
	\be
	\frac{\overline{\vert T_{ji}(t) \vert^2\vert T_{j'i}(t) \vert^2}}{ \overline{\vert T_{ji}(t) \vert^2} \;\overline{\vert T_{j'i}(t) \vert^2}}-1= \delta_{jj'}+C_2(t),
	\label{EqSpeckleCorrelationTime}
	\ee
	where $C_2(t)$ is a non-Gaussian contribution that is independent of $j,j'$ and therefore long-range. In Ref.~\cite{Skipetrov2004}, it is shown that in a quasi-1D system, and for $\tau_\text{Th} \lesssim t \lesssim \sqrt{g} \,\tau_\text{Th}$, $C_2(t)$ takes the simple form
	\be
	C_2(t)=\frac{2}{3g}\left(\alpha +\beta \frac{t}{\tau_{\text{Th}}} \right),
	\ee
	with $\alpha \propto 1/\sqrt{N_\nu}$ and $\beta \propto 1/N_\nu$ in the broadband limit $N_\nu\gg1$. This implies that for $t=t^* \sim \tau_{\text{Th}}$ [see Eq.~\eqref{EqArrivalTime}], we obtain
	\be
	C_2(t^*) \propto \frac{1}{g\sqrt{N_\nu}}.
	\ee
	Combining Eqs.~\eqref{EqDefNeffTime} and~\eqref{EqSpeckleCorrelationTime}, we then find
	\be
	\frac{1}{N_\text{eff}(t)}=\frac{1}{N_x} + C_2(t).
	\ee
	
	Assuming that $T(t)^\dagger T(t)$ behaves as a Wishart matrix with aspect ratio $N_1/N_\text{eff}(t)$, its eigenvalue distribution follows the Marchenko-Pastur law, whose upper edge satisfies the relation 
	\begin{align}
		\frac{\overline{\text{max}[\sigma(t)]}}{\overline{\sigma(t)}}&=\left[1+\sqrt{\frac{N_1}{N_\text{eff}(t)}} \right]^2
		\label{EqEnhancementTimeGated}
		\\
		&=\left[1+\sqrt{\frac{N_1}{N_x}+N_1C_2(t)} \right]^2,
	\end{align}
	with $\overline{\sigma(t)}=P_\text{un}(L,t) = \bar{T} \mathcal{F}(L,t)$ [see Eq.~\eqref{EqMeanIntensity}].
	This shows in particular that for $N_1=N_x$, and in the broadband limit, spatial-only shaping leads to the scaling
	\be
	\frac{P(L,t^*)}{P_\text{un}(L,t^*)} \simeq N_xC_2(t^*) \propto \frac{L}{\ell_t}\frac{1}{\sqrt{N_\nu}},
	\ee
	which is smaller by a factor $N_\nu$ compared to the result~\eqref{EqEnhancementCompleteControl} obtained for spatiotemporal control.
	
	In Fig. 2 of the main text, the theoretical prediction based on Eqs.~\eqref{EqEnhancementTimeGated} and~\eqref{EqDefNeffTime2} is shown to be in good agreement with simulation and experimental results.

    \section{Loading and Firing}
	\label{Sec:GTRvsTR}

\begin{figure*}
	\includegraphics[width=0.8\textwidth]{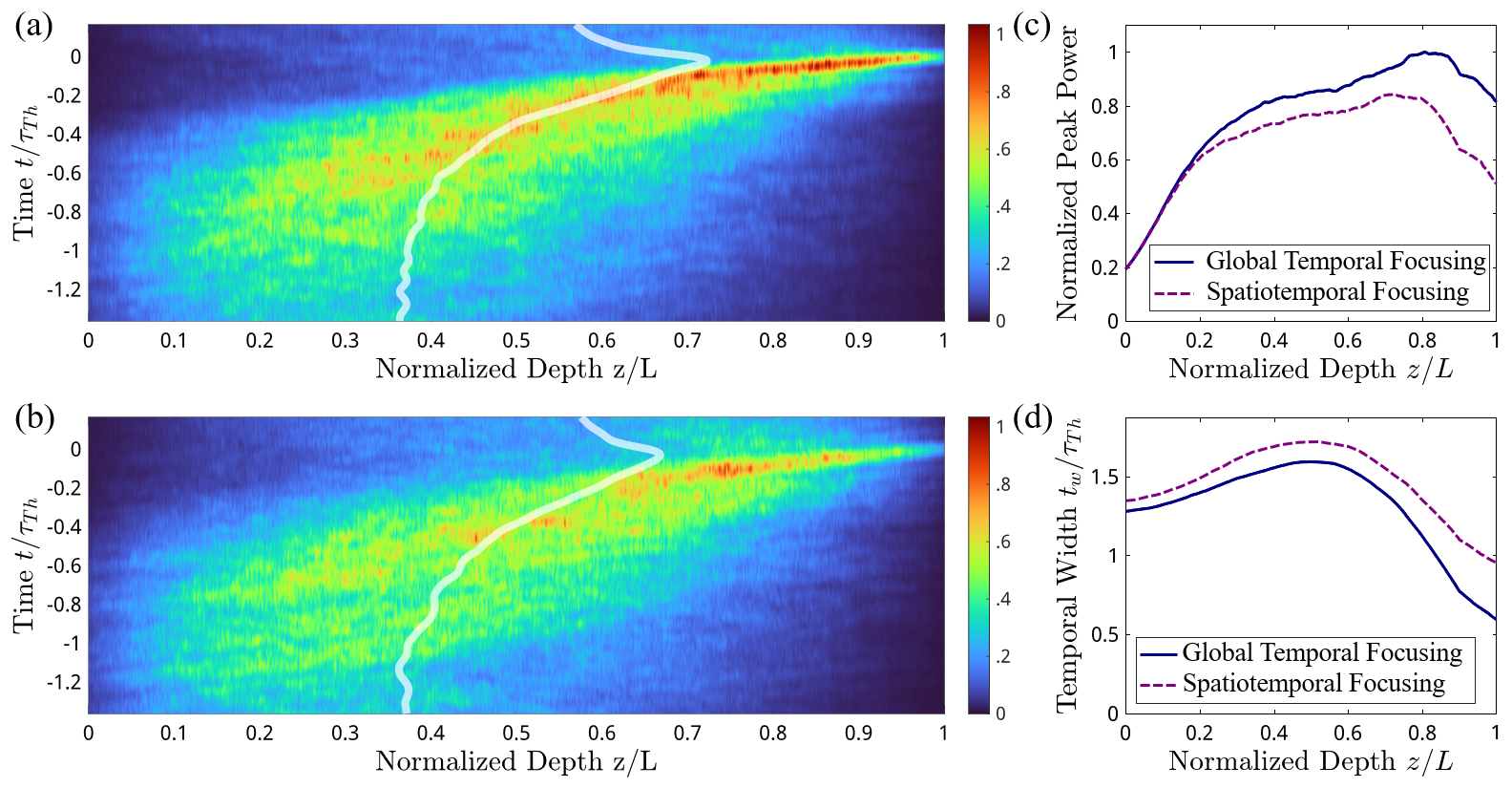}
	\caption{\label{FigFoc} \textbf{Global temporal focusing vs. spatiotemporal focusing.} Numerical simulation of a diffusive waveguide with $L$ = 100 \textmu m, $W$ = 30 \textmu m, $l_t$ = 3.3 \textmu m and no dissipation. (a) Cross-section integrated intensity profile $P(z,t)$ of the dominant spatiospectral channel propagating through a diffusive waveguide. The pulse with a bandwidth of $\Delta\nu = 42 \, \delta\nu$ is injected to the left end of the waveguide. Axial mean position of light intensity inside the disordered waveguide $\bar{z}(t)$ (white line) reveals the transition from loading to firing. (b) $P(z,t)$ for spatiotemporal focusing to a position $x_f$ at the right end $z=L$ of the waveguide. $\bar{z}(t)$ has a similar evolution with time but displays weaker firing. The time $t$ is normalized by the Thouless time $\tau_{\text{Th}} = 7.33$ ps and the depth $z$ is normalized by the waveguide length $L$. All color scales are normalized to the maximum value in (a). (c) Peak power as a function of depth for (a,b), normalized to the maximum value for (a). The peak power for the global temporal focusing (blue solid) is higher than the spatiotemporal focusing (purple dashed) at all depths. (d) Temporal pulse width $t_w(z)$ for the global temporal focusing (blue solid) is narrower than that of spatiotemporal focusing (purple dashed), particularly near the waveguide exit.}
\end{figure*}

Loading and firing dynamics appear when maximizing the peak transmitted power through a diffusive medium. We numerically identify a similar behavior when performing spatiotemporal focusing through the medium. According to time-reversal, spatiotemporal focusing corresponds to phase conjugating a single row of the tSSTM for a chosen spatial location at the output. To determine the pulse dynamics as a function of time and depth, we simulate both the dominant tSSTM eigenchannel and spatiotemporal focusing through a large diffusive waveguide of length $L = 100$ \textmu m, width $W = 30$ \textmu m, and transport mean free path $\ell_t = 3.3$ \textmu m. The resulting power distribution inside the waveguide is shown in Fig.~\ref{FigFoc}.

For global temporal focusing (temporal focusing to all spatial channels) via the dominant tSSTM eigenchannel, shown in Fig.~\ref{FigFoc}a, the same loading-and-firing dynamics are observed as in the main-text experiment in Fig.~\ref{Fig3}a. However, spatiotemporal focusing, Fig.~\ref{FigFoc}b, also produces a similar profile with a loading phase and a firing phase, although the total injected power is lower and the pulse firing is weaker. The mean axial positions of the pulses, $\bar{z}(t) = \int P(z,t) \, z \, dz /\int P(z,t) \, dz$ in Fig.~\ref{FigFoc}a and Fig.~\ref{FigFoc}b reveal a similar evolution with time, but firing is weaker for temporal focusing to a single spatial channel. Likewise, the peak power inside the sample is reduced at all depths (Fig.~\ref{FigFoc}c) and the temporal width, $t_w(z)$, is increased, especially at $z=L$  (Fig.~\ref{FigFoc}d). The temporal pulse width is defined by the participation number, $t_w(z) = \left[ \int P(z,t) \, dt \right]^2/\int \left[ P(z,t) \right]^2  \, dt$. The similarity between the two profiles is notable, because global temporal focusing is determined by long-range correlations while the spatiotemporal focusing enhancement should primarily depend on short-range correlations. However, even for spatial focusing of monochromatic light through a scattering medium, the effect of spatial long-range correlations on the background transmission enhancement is large~\cite{Shaughnessy2024}, increasing the total transmission to $2/3$ in the case of full spatial control~\cite{Vellekoop2008}. Therefore, the spatiotemporal dynamics inside the scattering medium for spatiotemporal focusing may also depend strongly on long-range correlations. To investigate this, we derive the power distribution inside the medium for spatiotemporal focusing, not only quantifying the importance of long-range correlations but also providing an analytic prediction for the spatiotemporal profile of the loading and firing dynamics.

For spatiotemporal focusing at position $z=L$ and time $t=0$, each frequency component of the input state is of the form $\ket{\psi_{\text{in}}(\nu)} = \mathcal{N} T(\nu)^\dagger \ket{x_f}$, where $\ket{x_f}$ represents an arbitrary focusing location, corresponding to the $f$th row of the transmission matrix $T(\nu)$. The state is normalized by the prefactor $\mathcal{N} = \bra{x_f} T(\nu) T(\nu)^\dagger \ket{x_f}^{-1/2} = \bar{T}^{-1/2}$. 
The resulting field $\ket{\psi(z,t)}$ at depth $z$ and time $t$ is then given by Eq.~\eqref{EqFieldDepthZ}, so that the mean power $P(z,t) = \overline{\sprod{\psi(z,t)}{\psi(z,t)}}$ takes the form
\begin{widetext}
	\begin{align}
		P(z,t)
		&= \iint_{\Delta \nu} \frac{d\nu_1}{\Delta \nu} \frac{d\nu_2}{\Delta \nu} \sum_{i,j,k}^{N_x} \frac{e^{-i2\pi(\nu_1-\nu_2)t}}{N_x \bar{T}}
		\overline{ \mathcal{Z}_{ji}(z,\nu_1) \mathcal{Z}_{jk}(z,\nu_2)^* T_{fk}(\nu_2) T_{fi}(\nu_1)^* }.
		\label{EqFocusingIntensity}
	\end{align}
\end{widetext}

\begin{figure*}[t]
	\includegraphics[width=0.75\textwidth]{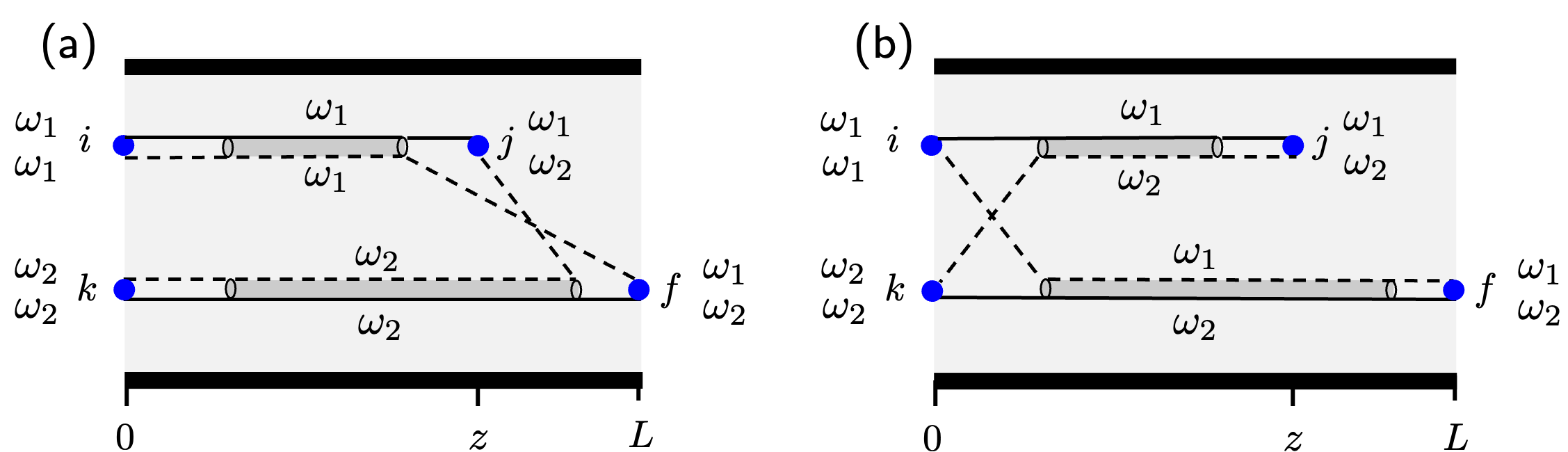}
	\caption{ {\bf Two Gaussian contributions to spatiotemporal focusing.} The four-field correlation function in Eq.~\eqref{EqFocusingIntensity} contains two Gaussian contributions. Panel (a) shows the contribution corresponding to $P_f(z,t)$ and panel (b)  the one corresponding to $P_s(z,t)$, see Eq.~\eqref{EqDecompoIntensity}. Their mathematical expressions are given by the first two terms of Eq.~\eqref{EqContraction}. Solid and dashed lines represent the fields and their complex conjugates, respectively; shaded tubes represent diffusive paths, and open circles indicate scatterers.
		\label{FigDiagramsGaussian}
	}
\end{figure*}

The average of the product of four fields in the integral can be split into three terms --- two Gaussian contributions and one non-Gaussian contribution:
\begin{widetext}
	\begin{align}
		&\overline{ \mathcal{Z}_{ji}(z,\nu_1) T_{fi}(\nu_1)^* } \; \overline{ \mathcal{Z}_{jk}(z,\nu_2)^* T_{fk}(\nu_2) }
		+ \overline{ \mathcal{Z}_{ji}(z,\nu_1) \mathcal{Z}_{jk}(z,\nu_2)^* } \; \overline{ T_{fk}(\nu_2) T_{fi}(\nu_1)^* }
		+ \overline{(\text{non-Gaussian})}
		\nonumber\\
		&= \delta_{z,L}\, \delta_{f,j} \frac{\bar{T}^2}{N_x^2} 
		+ \delta_{i,k} \frac{\bar{T} \bar{\zeta}(z)}{N_x^2} C_E(z,\omega_1-\omega_2) C_E(L,\omega_2-\omega_1) 
		+ \frac{\bar{T} \bar{\zeta}(z)}{N_x^2} \tilde{C}_2(z,\omega_1-\omega_2),
		\label{EqContraction}
	\end{align}
\end{widetext}
where $C_E(z,\omega_1-\omega_2)$ is the field correlation function already introduced in Sec.~\ref{Sec:Unmodulated}, and $\tilde{C}_2(z,\omega_1-\omega_2)$ is the non-Gaussian contribution discussed below. The diagrams corresponding to the two Gaussian contributions are shown in Fig.~\ref{FigDiagramsGaussian}.

Combining Eqs.~\eqref{EqFocusingIntensity} and~\eqref{EqContraction}, we find
\be
P(z,t) = P_f(z,t) + P_s(z,t) + P_l(z,t),
\label{EqDecompoIntensity}
\ee
where $P_f(z,t) = \bar{T}\, \text{sinc}^2(\pi \Delta \nu t)\, \delta_{z,L}$ is the main contribution at the focus $\ket{x_f}$, and $P_s(z,t)$ and $P_l(z,t)$ are the short- and long-range contributions, respectively:
\begin{widetext}
	\begin{align}
		P_s(z,t) &= 2\, \bar{\zeta}(z)\, \text{Re}\left[\int_{\Delta \omega} \frac{d\Omega}{\Delta\omega} \frac{\Delta\omega-\Omega}{\Delta\omega} e^{-i\Omega t} C_E(z,\Omega) C_E(L,-\Omega) \right],
		\label{EqIs}
		\\
		P_l(z,t) &= 2\,N_x \bar{\zeta}(z)\, \text{Re}\left[\int_{\Delta \omega} \frac{d\Omega}{\Delta\omega} \frac{\Delta\omega-\Omega}{\Delta\omega} e^{-i\Omega t} \tilde{C}_2(z,\Omega) \right].
		\label{EqIl}
	\end{align}
\end{widetext}
\begin{figure*}[t]
	\includegraphics[width=1\textwidth]{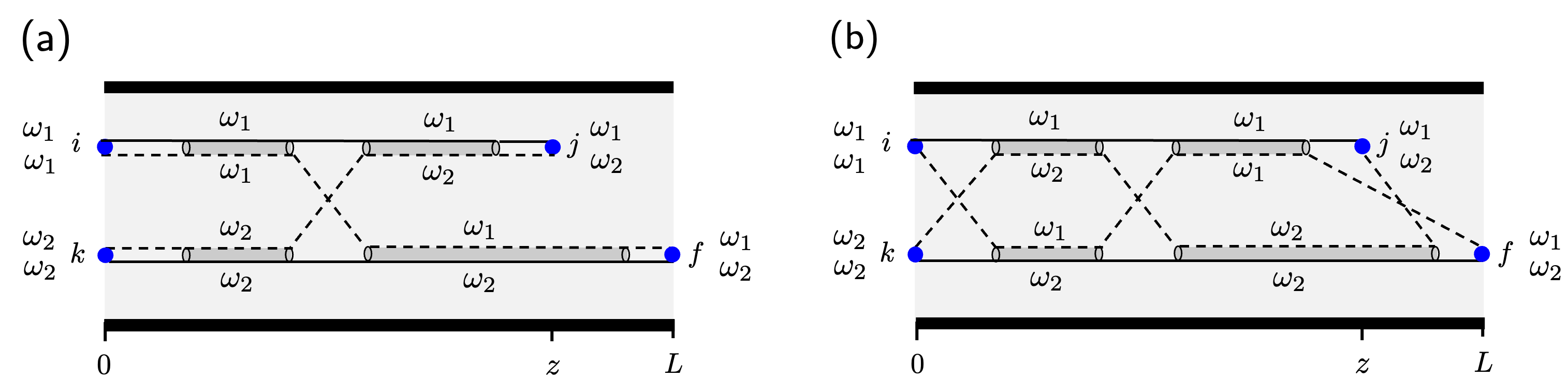}
	\caption{ {\bf Two contributions to the non-Gaussian correlation $\tilde{C}_2(z, \omega_1 - \omega_2)$ in spatiotemporal focusing.} Panel (a) shows the leading non-Gaussian contribution, Eq.~\eqref{EqC2FocusingLeading}, to the four-field correlation function $\overline{ \mathcal{Z}_{ji}(z,\nu_1) \mathcal{Z}_{jk}(z,\nu_2)^* T_{fk}(\nu_2) T_{fi}(\nu_1)^* }$ in Eq.~\eqref{EqFocusingIntensity}, while panel (b) shows a subleading non-Gaussian contribution. 
		The exchange of field partners between diffusive paths is a simplified depiction of a Hikami box. Summation over the indices $i$, $j$, and $k$ in Eq.~\eqref{EqFocusingIntensity} gives diagram (b) a subleading weight relative to (a), even when $z$ is close to the focusing depth $L$.
		\label{FigDiagrams}
	}
\end{figure*}

Figure~\ref{FigDiagrams}(a) shows the leading contribution to $\tilde{C}_2(z,\Omega)$, which arises from pairs of paths that propagate diffusively without dephasing before exchanging field partners inside the medium. These new partners then propagate with dephasing, due to the frequency difference $\Omega=\omega_1-\omega_2$, toward the focusing and detection planes located at $L$ and $z$, respectively.
The fact that dephasing occurs for paths connecting to the target, but not for those connecting to the input, reflects the strong variation of the power as the depth $z$ approaches $L$. This is directly related to the input state being designed to focus at $z = L$. Such a contribution can be written as
\begin{widetext}
	\be
	\tilde{C}_2(z,\Omega) = \frac{2}{gL\langle \mathcal{I}(z,0)\rangle \langle \mathcal{I}(L,0)\rangle} \int_0^L dz' \langle \mathcal{I}(z',0)\rangle^2 \partial_{z'}K(z,z',\Omega) \partial_{z'}K(L,z',-\Omega),
	\label{EqC2FocusingLeading}
	\ee
	where $g = N_x \bar{T}$ is the bare conductance of the disordered waveguide. In this expression, $\langle \mathcal{I}(z',0)\rangle^2$ represents the two non-dephasing diffusive paths that propagate up to the depth $z'$ where the field partner exchange occurs, while $K(z,z',\Omega)$ and $K(L,z',-\Omega)$ account for the dephasing paths that subsequently diffuse to the depths $z$ and $L$, respectively.  Explicit calculation for $\ell_t \ll L, L_\Omega$ and without absorption gives
	\be
	\tilde{C}_2(z,\Omega) = \frac{4}{g}\frac{1}{\tilde{L}(\tilde{L}-\tilde{z})}
	\frac{
		\left[1-\frac{i(\tilde{L}-\tilde{z})^2}{2} \right]
		\text{sinh}\left[\frac{(1+i)\tilde{z}}{2}\right]\text{sin}\left[\frac{(1+i)\tilde{L}}{2}\right]
		-\text{sinh}\left[\frac{(1+i)\tilde{L}}{2}\right]\text{sin}\left[\frac{(1+i)\tilde{z}}{2}\right]
	}
	{\text{cosh}(\tilde{L})-\text{cos}(\tilde{L})}.
	\label{EqC2TR}
	\ee
\end{widetext}
We stress that this long-range correlation function differs from the leading contribution to the intensity correlation function 
$\mathcal{C}(z,\Omega) = \overline{\mathcal{I}(z,\omega + \Omega/2) \, \mathcal{I}(L,\omega - \Omega/2)}/\overline{\mathcal{I}(z,\omega)} \, \overline{\mathcal{I}(L,\omega)} - 1$,
where $\mathcal{I}(z,\omega) = \sum_j |\mathcal{Z}_{ji}(z,\omega)|^2$ denotes the total intensity deposited at depth $z$ by exciting spatial channel $i$ at the waveguide entrance~\cite{Sarma2014}. The difference is due to the fact that dephasing now occurs before the partner exchange, in contrast to the long-range correlation of time-reversed waves given by Eq.~\eqref{EqC2FocusingLeading}, where dephasing occurs afterward [see also Fig.~\ref{FigDiagrams}(a)]. Both correlation functions coincide only at $\Omega = 0$, where $\tilde{C}_2(z,0) = (2/3g)\, z(2L - z)/L^2$.

\begin{figure}[h!]
	\includegraphics[width=0.5\textwidth]{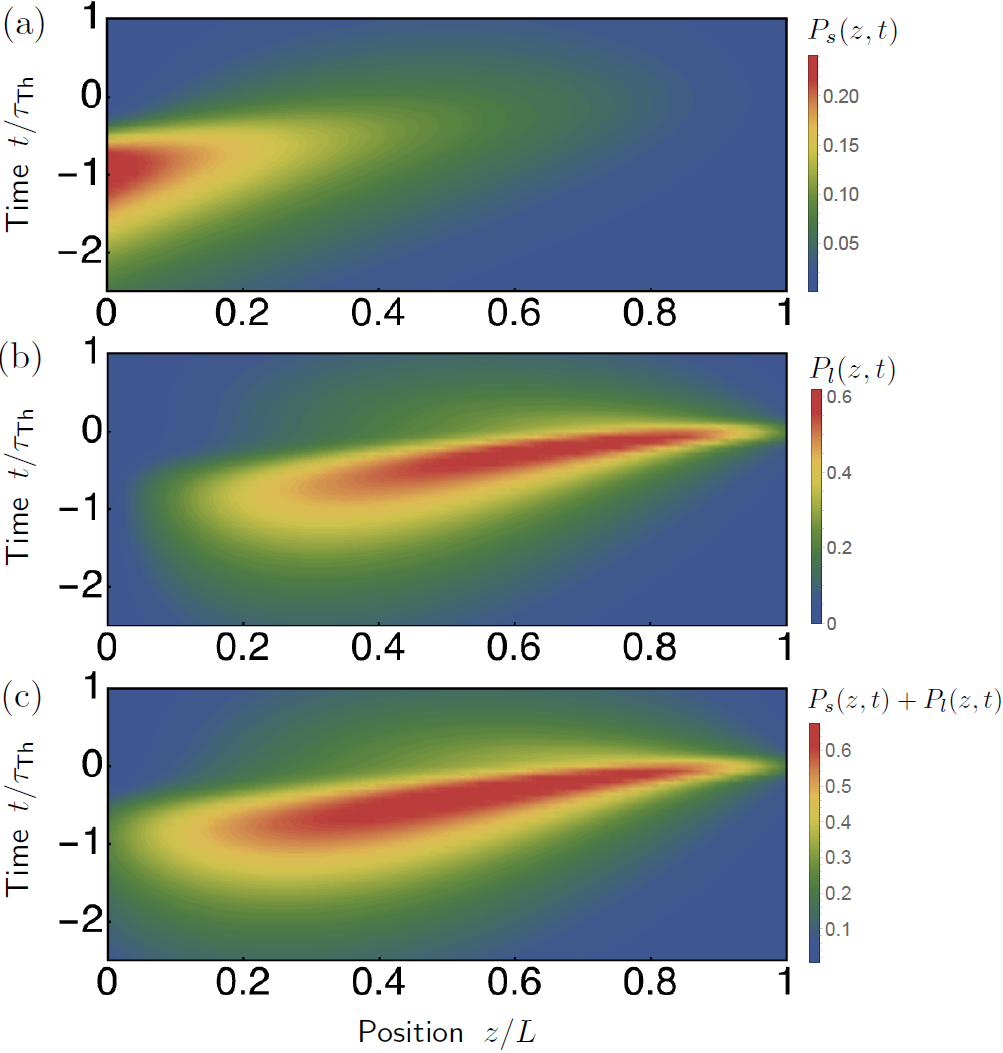}
	\caption{\label{FigSpatioTemporalFocusing_theory} 
		\textbf{Theoretical predictions for internal power distribution in spatiotemporal focusing.
		} The power of the propagating field for focusing at $z=L$ and $t=0$ [see Eqs.~\eqref{EqDecompoIntensity},~\eqref{EqIs}, and~\eqref{EqIl}] is decomposed into a short-range component, $P_s(z,t)$, located near the front surface (a), and a long-range component, $P_l(z,t)$, that extends deep inside the medium from the target (b), and dominates the total power $P(z,t)$ in (c). The parameters used for the calculation ($\bar{T}=0.1$ and $\Delta \omega/\omega_{\text{Th}}=28$), are chosen to be consistent with those of the experiment ($L=50\,\mu$m, $\ell_t=3.3\,\mu$m, $\Delta \nu/\delta\nu=10.5$).  
	}
\end{figure}

The theoretical prediction for the spatiotemporal profile $P(z,t)$ given by Eq.~\eqref{EqDecompoIntensity} is shown in Fig.~\ref{FigSpatioTemporalFocusing_theory}. The spectral bandwidth of the input signal and the mean transmission are chosen to be close to those of the experiment (see figure caption for details). The profile of $P_s(z,t)$ can be understood by noting that Eq.~\eqref{EqIs} can be expressed as the convolution, in the time domain, of the power at depth $z$ for a transform-limited pulse with unmodulated input wavefront, $P_\text{un}(z,t)$, with the time-reversal of that at depth $L$, $P_\text{un}(L,-t)$. Indeed, the comparison of Eqs.~\eqref{EqMeanIntensity} and~\eqref{EqDwellTime} with~\eqref{EqIs} shows that
\be
P_s(z,t) = \Delta \nu \, P_\text{un}(z,t) \otimes P_\text{un}(L,-t).
\ee
Since in the broadband limit the temporal profile $P_\text{un}(L,t)$ resembles a deformed pulse of width $\sim \tau_{\text{Th}}$, centered at time $t \sim \tau_{\text{Th}}$ (see Fig.~\ref{FigArrivalTime}), we find that $P_s(z,t)$ is concentrated near the front surface, like $P_\text{un}(z,t)$, but centered at time $t\sim-\tau_{\text{Th}}$ and with a temporal width on the order of the Thouless time. Like $P_\text{un}(z,t)$, $P_s(z,t)$ decays with depth $z$, so that its contribution to $P(z,t)$ can be neglected for $z/L \gtrsim 0.5$. In this part of the slab, $P(z,t)$ is dominated by $P_l(z,t)$. Contrary to $P_s(z,t)$, $P_l(z,t)$ increases with depth because it captures spatially long-range correlations between the depth $z$ and the focus at position $L$, which tend to increase as $z$ approaches $L$. Our predictions are in good agreement with the simulation of time-reversal focusing [see Fig.~\ref{FigFoc}(b)] and qualitatively reproduce the profile obtained with global temporal focusing [see Fig.~\ref{FigFoc}(a)] with the dominant tSSTM eigenchannel.  

\begin{figure*} [t]
	\includegraphics[width=0.7\textwidth]{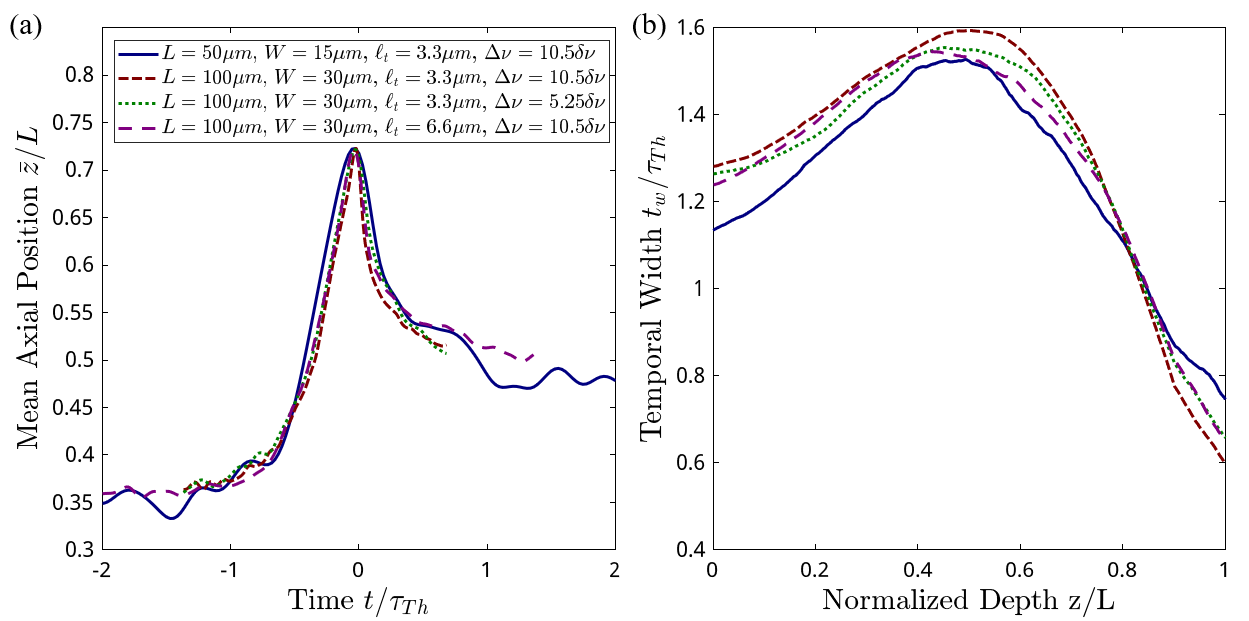}
	\caption{\label{FigParams} \textbf{Scaling of spatiotemporal profiles with sample parameters and input bandwidth.} Numerical simulation of global temporal focusing in four waveguides with varying length $L$, width $W$, transport mean free path $\ell_t$, and pulse bandwidth $\Delta\nu$. (a) Temporal evolution of the axial mean position of light intensity inside the disordered waveguide $\bar{z}(t)$ is invariant when the depth is normalized by $L$ and the time $t$ is normalized by the Thouless time $\tau_{\text{Th}}$. (b) Temporal pulse width $t_w(z)$ normalized by $\tau_{\text{Th}}$ has similar dependence on normalized depth $z/L$, except near the waveguide exit $z = L$ where the pulse width is determined solely by the spectral bandwidth $\Delta \nu$. The full-width-at-half-maximum (FWHM) of the transmitted pulse is approximately $1/\Delta\nu$.}
\end{figure*}

A close look at Eqs.~\eqref{EqIs},~\eqref{EqIl},~\eqref{EqCE}, and~\eqref{EqC2TR} reveals that $P(z,t)$ is parametrized by four quantities: the length $L$, the Thouless time $\tau_{\text{Th}}$, the mean transmission $\bar{T}$, and the bandwidth $\Delta \nu$. In the broadband limit where $P(z,t)$ is dominated by $P_l(z,t)$, $P(z,t)$ becomes proportional to the Fourier transform of the long range-function $\tilde{C}_2(z,\Omega)$ given by Eq.~\eqref{EqC2TR}, so that its spatial and temporal profile depend only on two scaling parameters, $L$ and $\tau_{\text{Th}}$, while $\bar{T}$ and $\Delta \nu$ affect only the global magnitude. However, this property is immediately apparent only in the case of spatiotemporal focusing to a single spatial location. 

We would like to investigate the scaling of $P(z,t)$ for global temporal focusing, which results from the propagation of the dominant eigenvector of the spatiospectral transmission matrix. To this end, we perform numerical simulations with four distinct sets of parameters. We start with a large diffusive waveguide (red dashed): $L = 100$ \textmu m, $W = 30$ \textmu m, $\ell_t = 3.3$ \textmu m, and $\Delta\nu = 10.5 \, \delta\nu$. In the subsequent simulations we halve the system dimensions (blue solid), half the input bandwidth (green dotted), and double the transport mean free path $\ell_t$ (purple long dashed). In each case, we calculate the axial mean position $\bar{z}(t) = \int P(z,t) \, z \, dz /\int P(z,t) \, dz$, Fig.~\ref{FigParams}a, and the temporal width $t_w(z) = \left[ \int P(z,t) \, dt \right]^2/\int \left[ P(z,t) \right]^2  \, dt$, Fig.~\ref{FigParams}b. For each simulation, the integral bounds for calculating $t_w(z)$ are $-1.36\tau_{\text{Th}}$ to $0.68\tau_{\text{Th}}$ rather than -10 to 5 ps as in Figs.~\ref{FigFoc} and~\ref{FigAbs}. This change is necessary because the definition of $t_w(z)$ strongly depends on the integral bounds, so they must be in units of $\tau_{\text{Th}}$ for a fair comparison. Plotting these results with depth normalized by $L$ and time by $\tau_{\text{Th}}$ shows good agreement between all four curves, with the exception of the fired pulse width $t_w(L)$, which primarily depends on the spectral width $\Delta \nu$. This suggests that the parameters $L$ and $\tau_{\text{Th}}$ determine the loading and firing dynamics $P(z,t)$ for global temporal focusing, similar to that predicted for spatiotemporal focusing.

\section{Effect of Dissipation}
\label{Sec:Abs}

\begin{figure*}
	\includegraphics[width=0.9\textwidth]{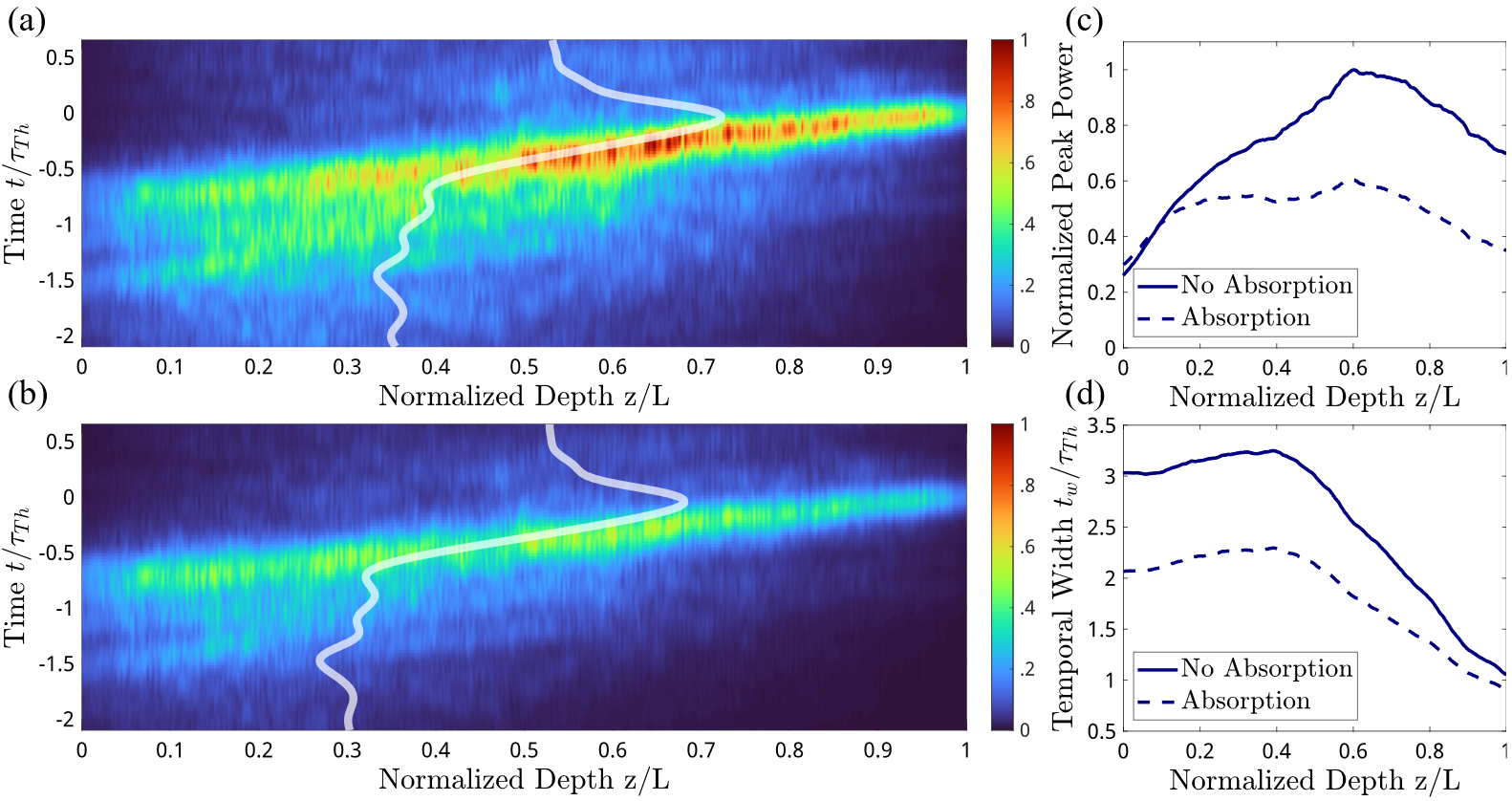}
	\caption{\label{FigAbs} \textbf{Effect of dissipation on pulse loading and firing in a diffusive waveguide.} Numerical simulation of a diffusive waveguide ($L$ = 50 \textmu m, $W$ = 15 \textmu m, $l_t$ = 3.3 \textmu m) without and with absorption ($\xi_a$ = 28 \textmu m). Cross-section integrated intensity profile $P(z,t)$ without dissipation (a) is compared to that with absorption (b). The pulse with a bandwidth of $\Delta\nu = 10.5 \delta\nu$ is injected to the left end of the waveguide. Axial mean position of light intensity inside the disordered waveguide $\bar{z}(t)$ (white line) reveals the loading to firing transition shifts slightly towards the waveguide entrance in the presence of absorption. The time $t$ is normalized by the Thouless time $\tau_{\text{Th}} = 1.83$ ps and the depth $z$ is normalized by the length $L$. All color scales are normalized to the maximum value in (a). (c) Peak power as a function of depth for (a,b), normalized to the maximum value for (a). The peak power without absorption (blue solid) is higher than that with absorption (blue dashed) at all depths. (d) Temporal pulse width $t_w(z)$ without absorption (blue solid) is wider than that with absorption (blue dashed), especially in the front half of the waveguide where the loading process dominates.}
\end{figure*}

Our experimental sample has dissipation due to out-of-plane scattering. To investigate the effect of dissipation on the loading and firing dynamics, we conduct numerical simulations of dominant tSSTM eigenchannels in a diffusive waveguide by introducing effective absorption to the system. The simulated waveguide has length $L = 50$ \textmu m, width $W = 15$ \textmu m, transport mean free path $\ell_t = 3.3$ \textmu m, and a diffusive dissipation length $\xi_a = 28$ \textmu m. $P(z,t)$ is shown without absorption in Fig.~\ref{FigAbs}a and with absorption in Fig.~\ref{FigAbs}b. The difference in the spatiotemporal profiles is quantified by calculating the axial mean position of the pulse $\bar{z}(t)$ inside the waveguide as a function of time. The results are plotted as white lines. A smaller $\bar{z}(t)$ in the loading stage reveals that the transition from loading to firing shifts slightly towards the waveguide entrance in the presence of absorption. Most notably, absorption narrows the temporal width throughout the waveguide, plotted in Fig.~\ref{FigAbs}d. This is particularly apparent in the front half of the sample where pulse loading occurs. Figure~\ref{FigAbs}d explicitly shows shortening of the loading phase by a 1.5-fold narrowing of the temporal width at the waveguide entrance. This is because absorption preferentially attenuates diffusive light with longer arrival times. As a result, the peak power throughout the waveguide is reduced compared to the case without absorption, shown in Fig.~\ref{FigAbs}c.

\section{Deep Pulse Injection}
\label{Sec:DPI}

\begin{figure*}
		\includegraphics[width=0.9\textwidth]{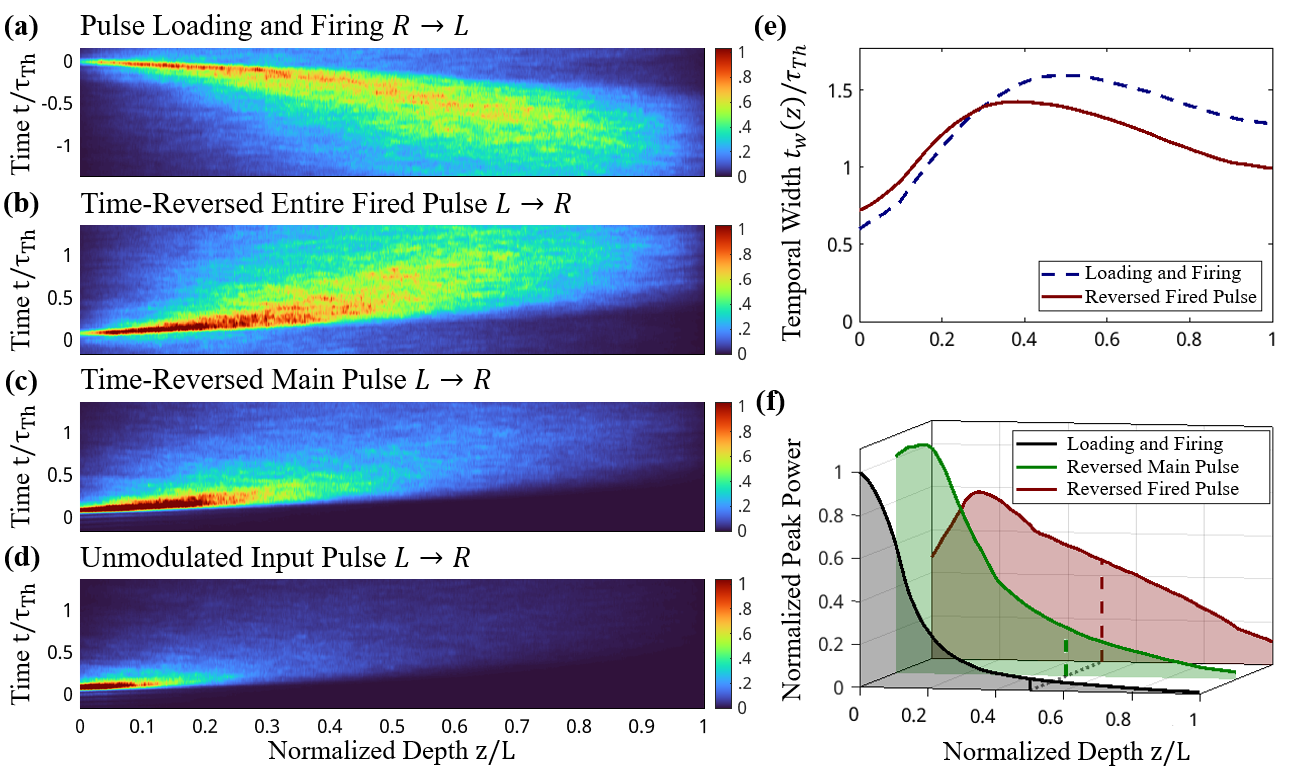}
		\caption{\label{FigDeepPulseSim} \textcolor{black}{\textbf{Pulse injection via time-reversed loading and firing.} (a) Numerically calculated spatiotemporal power distribution for the dominant tSSTM eigenchannel from right to left ($R\to L$), showing pulse loading and firing in a diffusive waveguide with length $L$ = 100 \textmu m, width $W$ = 30 \textmu m, transport mean free path $\ell_t$ = 3.3 \textmu m and no dissipation. The pulse bandwidth is $\Delta\nu = 42 \delta\nu$. (b) Time-reversing the transmitted field in (a) injects a pulse deep into the waveguide from the left end ($L\to R$). (c) Time-reversing only the spatial wavefront at $t = 0$ injects a transform-limited pulse into the sample ($L\to R$). (d) Launching a transform-limited pulse with an unmodulated wavefront leads to shallow injection ($L\to R$). The time $t$ is normalized by the Thouless time $\tau_\text{Th} = 7.33$ ps (average arrival time of light), and the depth $z$ normalized by the waveguide length $L = 100$ \textmu m. All color scales are normalized to the maximum value in (a). (e) Temporal pulse width $t_w(z)$ for loading and firing from right to left (blue dashed) in (a), and pulse injection via time-reversal from left to right (red solid) in (b), showing similar dependence on depth $z/L$. (f) Peak power in time as a function of depth $z/L$ for (b-d), normalized to the maximum for an unmodulated pulse. At $z/L = 1/2$, the peak powers in (b) and (c) are 9 times and 4 times of that in (d), respectively.}}
	\end{figure*}

    \textcolor{black}{In the main text, we experimentally reconstruct deep pulse injection in the dissipative waveguide. Since dissipation shortens the loading stage, time-reversed firing is not well separated from time-reversed loading in the waveguide of relatively small dimension. We thus perform numerical simulations of a larger waveguide without dissipation to better illustrate the time-reversed process of loading and firing.  By doubling the waveguide length and width and removing dissipation, the intrinsic dynamics of deep pulse injection become more apparent.}
    
    \textcolor{black}{We double the waveguide dimension to length $L=100$~\textmu m and width $W=30$~\textmu m. The transport mean free path remains the same $\ell_t=3.3$~\textmu m, and dissipation is removed. The tSSTM in the $R\to L$ direction provides the incident wavefront on the right that produces loading and firing toward the left end of the waveguide, as shown in Fig.~\ref{FigDeepPulseSim}a. We then time-reverse only the transmitted field at the left end and compute the resulting $L\to R$ spatiotemporal field evolution inside the waveguide.}
    
    \textcolor{black}{The internal power distribution $P(z,t)$ in Fig.~\ref{FigDeepPulseSim}b shows that the pulse injected from the left reaches approximately the center of the waveguide, equal to a travel distance of about $15\ell_t$. This corresponds to time-reversed firing, followed by time-reversed loading inside the medium. The temporal width of the injected pulse increases with depth for $z\lesssim L/2$, opposite to the pulse compression observed during firing (Fig.~\ref{FigDeepPulseSim}e). For comparison, a transform-limited pulse with an unmodulated plane-wavefront and identical spectral bandwidth remains much closer to the input side, as shown in Fig.~\ref{FigDeepPulseSim}d.}
    
    \textcolor{black}{The transmitted pulse created by loading and firing contains a broad temporal background. Its time-reverse therefore injects light outside the main pulse with a time-varying spatial wavefront. To isolate the contribution of the main pulse from that of the background, we inject a transform-limited pulse with a spatial wavefront equal to the phase-conjugated field of the main pulse transmitted at $t=0$. According to gTR, this spatial wavefront is identical to the dominant eigenvector of the BTM in the $L\to R$ direction. The spatiotemporal power profile in Fig.~\ref{FigDeepPulseSim}c shows deeper pulse injection than the transform-limited pulse with unmodulated plane-wavefront in Fig.~\ref{FigDeepPulseSim}d, but it does not reach the depth achieved by time-reversing the entire fired pulse including the background in Fig.~\ref{FigDeepPulseSim}b.}
    
    \textcolor{black}{For a quantitative comparison, we plot the peak power as a function of depth in Fig.~\ref{FigDeepPulseSim}f. The peak power delivered by the input pulse with an unmodulated wavefront decays rapidly with depth. Switching the input wavefront to the dominant BTM eigenvector increases the peak power by a factor of 4 at $z=L/2$, while time-reversal of the entire fired pulse, including its temporal background, provides a 9-fold enhancement at the same depth. The latter also enhances the peak power at depths beyond $L/2$. 
    Supplementary Videos 3 and 4 show the numerical simulations of pulse loading and firing and deep pulse injection, respectively.}
	
	
\end{document}